\documentclass[reprint,showpacs,preprintnumbers,superscriptaddress,multibbl,amsmath,amssymb,prl]{revtex4-2}

\usepackage[utf8]{inputenc}
\usepackage{graphicx,wasysym}
\usepackage{dcolumn}
\usepackage{bm}
\usepackage[colorinlistoftodos, color=green!40, prependcaption]{todonotes}
\usepackage[pdftex, pdftitle={Article}, pdfauthor={Author}]{hyperref} 
\usepackage{amsthm}
\usepackage{mathtools}
\usepackage{physics}
\usepackage{xcolor}
\usepackage{graphicx}
\usepackage[left=23mm,right=13mm,top=35mm,columnsep=15pt]{geometry} 
\usepackage{adjustbox}
\usepackage{placeins}
\usepackage[T1]{fontenc}
\usepackage{lipsum}
\usepackage{csquotes}
\usepackage[english]{babel}
\begin{document}
\title{Noise driven phase transitions in eco-evolutionary systems}

    \author{Jim Wu}
    \affiliation{Department of Physics, Princeton University, Princeton, NJ 08544, USA}
    \affiliation{Center for the Physics of Biological Function, Princeton University, Princeton, NJ 08544, USA}
    \author{David J. Schwab}
       \email[Correspondence email address: ]{davidjschwab@gmail.com}
     \affiliation{Center for the Physics of Biological Function, Princeton University, Princeton, NJ 08544, USA}
    \affiliation{Initiative for the Theoretical Sciences, The Graduate Center, CUNY, New York, NY 10016, USA}
    \author{Trevor GrandPre}
    \email[Correspondence email address: ]{tgrandpr@princeton.com}
    \affiliation{Department of Physics, Princeton University, Princeton, NJ 08544, USA}
    \affiliation{Center for the Physics of Biological Function, Princeton University, Princeton, NJ 08544, USA}
    \affiliation{Princeton Center for Theoretical Science, Princeton University, Princeton, NJ 08544, USA}

\date{\today} 

\begin{abstract}
In complex ecosystems such as microbial communities, there is constant ecological and evolutionary feedback between the residing species and the environment occurring on concurrent timescales. Species respond and adapt to their surroundings by modifying their phenotypic traits, which in turn alters their environment and the resources available.
To study this interplay between ecological and evolutionary mechanisms, we develop a consumer-resource model that incorporates phenotypic mutations. In the absence of noise, we find that phase transitions require finely-tuned interaction kernels. Additionally, we quantify the effects of noise on frequency dependent selection by defining a time-integrated mutation current, which accounts for the rate at which mutations and speciation occurs. We find three distinct phases: homogeneous, patterned, and patterned traveling waves. The last phase represents one way in which co-evolution of species can happen in a fluctuating environment. Our results highlight the principal roles that noise and non-reciprocal interactions between resources and consumers play in phase transitions within eco-evolutionary systems.
\end{abstract}

\setcounter{secnumdepth}{6}
\maketitle

\section{Introduction}
Canonical ecological models construct a community of monomorphic species and explores how coexistence is shaped by the interactions with each other and with their environment \cite{lotkaAnalyticalNoteCertain1920,lotkaElementsMathematicalBiology1970,volterraVariazioniFluttuazioniNumero1926,volterraFluctuationsAbundanceSpecies1926,macarthurLimitingSimilarityConvergence1967,macarthurSpeciesPackingCompetitive1970,chessonMacArthurConsumerresourceModel1990}. These models generally overlook the evolutionary effects, assuming a static community of existing species and neglecting the changes in traits over time. In contrast, evolutionary models focus on how the distribution of traits in the community change over time while neglecting ecological interactions, absorbing these complexities within fitness differences \cite{kimuraStochasticProcessesDistribution1955,gillespieGeneticDriftInfinite2000,neherProgressOpenProblems2018}. At the core of both classes of models is the assumption that ecological and evolutionary processes typically occur on disparate timescales, especially in ecosystems of macroscopic organisms \cite{neherProgressOpenProblems2018}. Whereas birth, death, and competition tend to play out on fast timescales, new mutant species arise and fix within the community over longer timescales. As a result, populations settle to a steady state before a new mutant triggers a reorganization of the community structure \cite{doebeliAdaptiveDiversification2011}. However, in many cases there is no separation of time-scales and the coupling between ecology and evolution cannot be ignored. Understanding how ecology influences and feeds back into evolutionary dynamics, and vice versa, is an open problem and represents a frontier in the field. 

Within microbial ecosystems, the assumption of timescale separation breaks down especially for microbial ecosystems where mutations are more frequent~\cite{neherProgressOpenProblems2018,billerProchlorococcusStructureFunction2015,braakmanMetabolicEvolutionSelforganization2017,rosenFinescaleDiversityExtensive2015}.~In addition, there is mounting evidence of eco-evolutionary feedback among macroscopic ecosystems such as butterflies~\cite{hanskiMolecularLevelVariationAffects2006,zhengModellingSingleNucleotide2009}, zooplankton~\cite{palkovacsEcoevolutionaryInteractionsPredators2008,palkovacsExperimentalEvidenceThat2009,postIntraspecificVariationPredator2008}, plants~\cite{schweitzerGenesEcosystemsGenetic2008,schweitzerPlantsoilmicroorganismInteractionsHeritable2008}, and birds~\cite{palkovacsEcoevolutionaryDynamicsIntertwining2010,hendryCritiqueEcoEvolutionary2019,bailey2006importance}.

One of the early attempts to unite ecology and evolution into a single framework was adaptive dynamics, which assumes mutations are small and infrequent and natural selection acts quickly to propagate the fittest species~\cite{doebeliAdaptiveDiversification2011}. With its ability to capture the emergence of ecological diversity, adaptive dynamics has been applied to a variety of ecosystem models from ones with competitive interactions, cooperative interactions, and sexual reproduction~\cite{doebeliAdaptiveDiversification2011,doebeliDiversityCoevolutionaryDynamics2017,dieckmannDynamicalTheoryCoevolution}. Since adaptive dynamics assumes a separation of timescales, new theory was needed to incorporate concurrent ecological and evolutionary timescales. As a result, generalized Lotka-Volterra (gLV) models have been developed incorporating mutations using a diffusion process in niche space and demographic noise. These models have been shown to exhibit niche space patterning and traveling wave solutions \cite{rogersDemographicNoiseCan2012,rogersModesCompetitionFitness2015,rogersSpontaneousGeneticClustering2012,shnerbPatternFormationNonlocal2004}. In the context of ecology and evolution, the latter scenario embodies a coevolutionary arms race where species constantly adapt and evolve to survive against other ever-evolving competitors. Species at the front of a wave experience a competition asymmetry as there are no interacting species in front. As a result, these pioneering species grow rapidly and their large numbers inhibit the growth of the species behind. This feedback creates a depletion zone in the wake of the moving front, forming an oscillatory traveling wave in trait space. If the population of these depleted species dip below the genetic drift barrier, then they will likely go extinct and the pioneering species will split off from the group. This represents a diversification event and would explain how competition can give rise to sympatric speciation \cite{rogersDemographicNoiseCan2012,rogersModesCompetitionFitness2015,rogersSpontaneousGeneticClustering2012,shnerbPatternFormationNonlocal2004}.
	\begin{figure*}[t]
\begin{center}
\includegraphics[width=14cm]{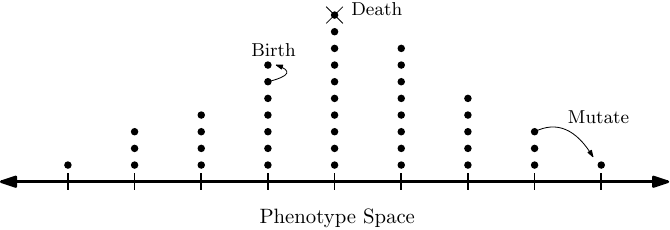}
\caption{ Consumer-resource model with mutations. Each tick mark is a specific phenotype. Each phenotype has an associated consumer  and resources number density. Each dot at each tick mark in phenotype space represents an individual consumer with a preference for a particular resource. From consuming and converting the available resources into biomass, the consumers give birth to new individuals. In addition, consumers can die and mutate to prefer other resource types. Since birth, death, and mutations occur randomly, the consumer-resource model intrinsically has demographic noise.}
\label{Fi:1}
\end{center} 
\end{figure*}

In addition to traveling waves, the nonlocal gLV equation also displays Turing patterns under conditions that largely depend on the form of the competition kernel and the nature of the noise. For a large class of competition kernels, a community of infinite size has a stable homogeneous solution. Any perturbations that induce heterogeneities in abundances will eventually equilibrate. However, if the competition kernel has negative Fourier modes, then the homogeneous state destabilizes and the abundance distribution exhibits a periodic Turing pattern in phenotype space~\cite{rogersDemographicNoiseCan2012,rogersModesCompetitionFitness2015,rogersSpontaneousGeneticClustering2012,shnerbPatternFormationNonlocal2004,pigolottiSpeciesClusteringCompetitive2007,pigolottiHowGaussianCompetition2010,hernandez-garciaSpeciesCompetitionCoexistence2009,lopezFluctuationsImpactPatternforming2004a,fortClumpingTransitionNiche2010}. The phenotypic clustering of species provides an explanation of how to maintain diversity in an evolving ecosystem, but requires finely tuned competition kernels to have negative Fourier modes. This model suggests that the emergence of diversity of ecosystems is highly dependent on the form of the competition kernels~\cite{rogersDemographicNoiseCan2012,pigolottiHowGaussianCompetition2010,pigolottiSpeciesClusteringCompetitive2007,fortClumpingTransitionNiche2010}. To remedy the lack of robustness in phentoypic pattern formation, some have included demographic noise into the models.  Noise is often thought to destroy ordered phenomena. In the case of Turing patterns, stochasticity has been reported to actually extend the range of parameters occupied by the Turing phase and make the patterns more robust \cite{biancalaniFrameworkAnalyzingEcological2015,biancalaniGiantAmplificationNoise2017,butlerFluctuationdrivenTuringPatterns2011a,shihPathintegralCalculationEmergence2014}. Hence, the noisy generalized Lotka-Volterra model with mutations provides a simple, effective theory of how competitive interactions and demographic noise promote ecological diversity. However, in real ecological systems, competition between species is mediated through the resources in the environment. This is not explicitly considered in gLV models. 

Thus, in this work, we develop a phenotypic model of microbial ecosystems based on the classic MacArthur consumer-resource model~\cite{macarthur1970species}. Our model generalizes the original consumer-resource model by incoporating consumer mutations,  and consumer and resource demographic noises, which accounts for evolutionary diversification and the inherent stochasticity of finite populations. Using large deviation theory to analyze the consumer-resource dynamics, we find three distinct phases: homogeneous, patterned, and patterned traveling waves. We define a speciation current, the rate that new species are created due to mutations, as a dynamical observable to distinguish whether an ecosystem stays in a homogeneous profile in niche space, spontaneously diversifies into distinct phenotypic clusters, or undergoes a coevolutionary arms race. We demonstrate that the onset of these robust species clusters in niche space is not predicated on the existence of negative Fourier modes in the interaction kernels, but rather on the demographic noise, the efficiency of converting resources into biomass, and the relative sizes of the consumer and resource trophic levels. This highlights the fact that niche differentiation is a robust process across many real ecosystems.

\section{Consumer-Resource Model}
Consider an $M$-dimensional lattice with each site ${\bf c}$ representing the phenotype of a particular species or resource. Each resource of trait ${\bf c}$ with density $r({\bf c})$ grows at a rate $gr({\bf c})$, and a decays at a rate $ gr^2({\bf c})/\kappa$. Here, $\kappa$ is the resource carrying capacity and $g$ is the growth rate constant. Consumers are also defined on the same phenotype space with $\rho({\bf c})$ representing the consumer density. 
Consumers of type ${\bf c}$ die at a rate of $\gamma \rho({\bf c})$ and 
consume resources at a rate of $Z_2\sum_{\bf c}\phi({\bf c}-{\bf c}') \rho({\bf c}')$, where $Z_2$ is the prefactor rate of consumption, $\phi({\bf c}-{\bf c}')$ is the kernal of consumer preference for resource type ${\bf c}'$, and $\gamma$ is the death rate constant.
However, not all of the consumed resources are used to generate new consumer biomass or produce offspring. For example, some of the energy is used to maintain the vital biological functions of the consumers. Therefore, consumer birth is modeled as $Z_1\sum_{\bf c} \psi({\bf c}-{\bf c}') r({\bf c}')$. Here, $\psi({\bf c}-{\bf c}')$ is defined as the efficiency of species ${\bf c}$ in converting resource ${\bf c}'$ into the biomass, and $Z_1$ is the prefactor rate of resource to biomass conversion. We define the ratios for the biomass conversion efficiency and relative consumer-resources abundance to be $\epsilon= Z_1/Z_2$ and $\chi=\kappa/N_c$, where $N_c$ is the total number of consumers in the system.

We make the assumption that the $\phi$ and $\psi$ kernels only depend on the phenotypic distance between the resource and consumer trait. A small phenotypic distance indicates a close match between the resource type and a species preference, whereas a large distance means that a species is not likely to consume that resource. In addition to birth, death, and consumption, species can also mutate at a rate $D(\rho)$ to nearby phenotypes. Assuming that the mutational step from ${\bf c}$ to ${\bf c}'$ is small, we can approximate the mutation process as a diffusion in trait space. In Fig. \ref{Fi:1}, a one-dimensional illustration of our model is shown.
In Appendix \ref{A1}, we define our model in further detail and derive our continuous-space model from a Van Kampen expansion of a discrete state model (see Appendix \ref{A1}):
\begin{align}
\label{eq:species}
	\frac{d\rho({\bf c})}{dt} &= A(\rho({\bf c}),r({\bf c})) +  \nabla\cdot\left( D(\rho) \nabla\rho({\bf c})\right)\\
 &\phantom{==}+ \sqrt{2\sigma_\rho(\rho)}\eta_\rho(t) + \nabla\left(\sqrt{2\sigma_d(\rho)}\eta_j(t) \right)\notag~,
 \end{align}
\begin{align}
	\label{eq:resources}
	\frac{dr({\bf c})}{dt} &= B(\rho({\bf c}),r({\bf c})) + \sqrt{2\sigma_r(r)}\eta_r(t)~,
 \end{align}
\begin{equation}
\label{Arho}
      \frac{A(\rho({\bf c}),r({\bf c}))}{\rho({\bf c})} =Z_1\int \psi({\bf c}-{\bf c}') r({\bf c}') d{\bf c}' - \gamma,
  \end{equation}
  and
  \begin{equation}
  \label{eq:resources3}
     \frac{B(\rho({\bf c}),r({\bf c}))}{r({\bf c})}=g - \frac{gr({\bf c})}{\kappa} - Z_2\int \phi({\bf c}-{\bf c}') \rho({\bf c}') d{\bf c}'~.
  \end{equation}
Here, we define the function $A(\rho,r)$ in Eq. \eqref{Arho} to contain the conversion of consumed resources into biomass and the death rate of consumers. Similarly, $B(\rho,r)$ in Eq. \eqref{eq:resources3} captures growth and decay as well as consumption of resources by the consumers. The terms $\sqrt{2\sigma_\rho(\rho)} \eta_\rho$, $\sqrt{2 \sigma_r(r)}\eta_r$, and $\sqrt{2\sigma_d(\rho)}\eta_j$ in Eqs. \eqref{eq:species} and \eqref{eq:resources} represent the intrinsic birth-death noise in the consumer and resource population and the noise in the mutation process, respectively. The noise terms $\eta_\rho$, $\eta_r$, and $\eta_j$ are modeled using Wiener processes and the multiplicative noise strengths, 
\begin{equation}
   \sigma_\rho(\rho)=\frac{\rho({\bf c})}{N_c} \left[Z_1\int \psi({\bf c}-{\bf c}') r({\bf c}') d{\bf c}' + \gamma\right]~,
\end{equation}
and
\begin{equation}
 \sigma_r(r)=\frac{r({\bf c})}{\kappa }\left[ g + \frac{gr({\bf c})}{\kappa} +Z_2 \int \phi({\bf c}-{\bf c}') \rho({\bf c}') d{\bf c}'  \right]
\end{equation}
 are derived using Poisson statistics (see Appendix \ref{A1}). Consequently, $\sigma_\rho$ and $\sigma_r$ are inversely proportional to the total consumer and resource abundances, $N_c$ and $\kappa$, respectively. 
 \begin{table*}[t]
\begin{center}
 \label{table1}
\scalebox{1.1}{\begin{tabular}{ccc}
		Process & Reaction & Model\\
		\hline
		\hline
		Resource growth & $R({\bf c}) \rightarrow 2R({\bf c})$  & $g r({\bf c})$\\
		Resource decay & $R({\bf c}) \rightarrow \emptyset$  & $gr^2({\bf c})/\kappa $\\
		Consumption & $N({\bf c}') + R({\bf c}) \rightarrow N({\bf c})$ & $Z_2\int\left[\phi({\bf c}-{\bf c}') \rho({\bf c}')d{{\bf c}'}\right]r({\bf c})$\\
		Birth &	$N({\bf c})\rightarrow 2 N({\bf c})$ & $Z_1 \int \left[\psi({\bf c}-{\bf c}') r({\bf c}')d{{\bf c}'}\right]\rho({\bf c}) $ \\
		Mutation & $N({\bf c}) \rightarrow  N({\bf c}')$ &  $\nabla\cdot(D(\rho) \nabla \rho({\bf c}))$ \\
		Death & $N({\bf c}) \rightarrow \emptyset$ & $\gamma \rho({\bf c})$
	\end{tabular}}
 \end{center}
\caption{ Summary of ecological and evolutionary processes in the consumer-resource model. The Reaction column refers to the lattice in Fig. 1 with dynamics defined in Appendix \ref{A1}, and the Model column refers to the continous-space model in Eqs. \eqref{eq:resources}-\eqref{eq:resources3}. The variables $R(\bf{c})$ and $N(\bf{c})$ are the resource and consumer population size, respectively. }
\end{table*}

The noise term $\sigma_d(\rho)$ couples diffusion of mutations to the density of a phenotype, $\rho({\bf c})$. In this work, we consider two general forms: free diffusion and excluded volume interactions. For free diffusion of particles on a lattice, the diffusion noise takes the form of  $\sigma_d(\rho)=D\rho$. In this case, the diffusion of mutations is independent of consumer-consumer and consumer-resources interactions. The other form we will consider is an excluded volume interaction, $\sigma_d(\rho)=D\rho\left(1-\rho\right)$. 
This form implies that that there is a reduction in the diffusion magnitude at higher consumer densities~\cite{agranovEntropyProductionIts2022a,agranovExactFluctuatingHydrodynamics2021,agranovMacroscopicFluctuationTheory2022,agranovTricriticalBehaviorDynamical2023,baekDynamicalPhaseTransitions2018,baekDynamicalSymmetryBreaking2017,bodineauCurrentLargeDeviations2006,bodineauDistributionCurrentNonequilibrium2005,appert-rollandUniversalCumulantsCurrent2008}. In our model, new mutants have difficulty establishing themselves if their traits strongly resemble those of a highly populous competitor. Furthermore, this form of the $\sigma_d(\rho)$ noise could originate from consumers freely mutating in both directions in niche space with a negative mutation-rate plasticity where a denser consumer population experiencing stronger competition has a lower mutation rate. Mounting experimental evidence shows a strong negative relation between mutation rate and population density within and across many species of prokaryotic and eukaryotic micro-organisms \cite{krasovecSpontaneousMutationRate2017,aanenMutationratePlasticityGermline2019}. In subsequent sections, we will study how noise gives rise to complex dynamical behavior. First, we conduct an analysis without noise terms to understand the the importance of noise on pattern formation. 
 \section{Results}
\subsection{Turing analysis: niche differentiation}

To understand the importance of noise, we first look at the deterministic dynamics in our model by setting the deterministic non-diffusional parts of Eqs.~\eqref{eq:species} and \eqref{eq:resources} to zero (see Appendix \ref{A2}). We find that 
	\begin{align}
		\rho^* &= \left(Z_2\int \phi({\bf c}-{\bf c}')~d{\bf c}'\right)^{-1}\left(1 - \frac{r_*}{\kappa}\right) \\&= \frac{1}{Z_2\tilde{\phi}(0)}\left(1 - \frac{\gamma}{Z_1 \tilde{\psi}(0) \kappa} \right)\label{eq:equil_rho}\\
		r_* &= \left(Z_1\int \psi({\bf c}-{\bf c}')~d{\bf c}'\right)^{-1} \gamma  = \frac{\gamma}{Z_1 \tilde{\psi}(0)} \label{eq:equil_r}
	\end{align}
	where $\tilde{\psi}(0)$ and $\tilde{\phi}(0)$ are the Fourier transforms of the kernels $\psi({\bf c})$ and $\phi({\bf c})$ evaluated at wavenumber ${\bf q} = 0$.  To probe the linear stability of the fixed point, we expand Eqs.~\eqref{eq:species} and \eqref{eq:resources} without noise around $\rho^*$ and $r^*$ respectively. This yields
	\begin{align}
		\frac{d\delta\rho({\bf c})}{dt} &=  Z_1\rho^*\int \psi({\bf c}-{\bf c}')\delta r({\bf c}')~d{\bf c}' + m\nabla^2 \delta \rho({\bf c}) \\
		\frac{d\delta r({\bf c})}{dt} &= r^*\left[- \frac{\delta r({\bf c})}{\kappa} -Z_2\int \phi({\bf c}-{\bf c}') \delta \rho({\bf c}') ~d{\bf c}'\right]
	\end{align}
	where we define $\delta \rho = \rho - \rho^*$ and $\delta r = r - r^*$ as the fluctuations in the consumer and resource densities. Performing the Fourier transform on these system of PDEs allows us to rewrite them in matrix form as $\omega \boldsymbol{\zeta} = {\bf J}\boldsymbol{\zeta}$ \cite{crossPatternFormationOutside1993,jiangTraitspacePatterningRole2019a} where 
 \begin{equation}
     {\bf J}= \begin{pmatrix}
		-D{\bf q}^2 & Z_1\rho^* \tilde{\psi}({\bf q}) \\
		-Z_2r^* \tilde{\phi}({\bf q})& -r^*/\kappa
		\end{pmatrix}~,
 \end{equation}
 and
 \begin{equation}
     \boldsymbol{\zeta}=\begin{pmatrix}
		\delta \tilde{\rho}({\bf q})\\ 
		\delta \tilde{r}({\bf q})	\end{pmatrix}~.
 \end{equation}
 Here, the diffusion is assumed to be a constant $D(\rho) = D$.  The eigenvalues of the matrix {\bf J} are found by solving the characteristic equation $\omega^2 - \text{tr}({\bf J})\omega + \text{det}({\bf J}) = 0$, from which we obtain the dispersion relation
	\begin{align}
		\omega({\bf q}) &= \frac{\text{tr}({\bf J}) \pm \sqrt{\left(\text{tr}({\bf J})\right)^2 - 4\cdot\text{det}({\bf J})}}{2}~.
	\end{align}
This suggests that the homogeneous state is stable if $\text{Re}[\omega({\bf q})] < 0$ for all ${\bf q}$. Conversely, if there exists at least one mode ${\bf q}_c$ where $\text{Re}[\omega({\bf q}_c)] > 0$, then homogeneous state is linearly unstable. As a result, the mode ${\bf q}_c$ grows, giving rise to a dynamical phase transition with a dispersion relation given by $\omega({\bf q}_c)$. This Turing pattern condition is satisfied only if $\text{det}({\bf J}) < 0$, or
	\begin{align}
            \frac{D}{\rho^*\kappa} < - \frac{\tilde{\psi}({\bf q}_c)\tilde{\phi}({\bf q}_c)}{{\bf q}_c^2}~.
		\label{eq:Turing_cond}
	\end{align}
	Since $D/(\rho^* \kappa)$ and ${\bf q}_c^2$ are both positive quantities, the inequality \eqref{eq:Turing_cond} can only be satisfied if the Fourier transforms of the biomass conversion kernel $\psi$ and consumption kernel $\phi$ are \emph{not} positive definite. In the scenario without mutations, $D = 0$, then the onset of Turing instability occurs at $\tilde{\psi}({\bf q})\tilde{\phi}({\bf q}) < 0$. This suggests that the interaction between consumers and resources is the primary driver of pattern formation rather than mutations.
	
These stringent conditions make the consumer-resource dynamics highly sensitive to the exact form of the $\psi$ and $\phi$ kernels. Gaussian and exponential kernels do not satisfy the Turing instability condition whereas tophat kernels and exponential of quartics can lead to pattern formation.  Furthermore, if the biomass conversion kernel is proportional to resource consumption through a power law $\psi({\bf q}) \propto \phi^\alpha({\bf q})$ with odd $\alpha$, then the inequality \eqref{eq:Turing_cond} is also never satisfied and pattern formation is impossible. The fine tuning of the consumer-resource interaction kernels to generate trait space patterns is rather unrealistic and provides an unsatisfactory explanation for diversification and speciation in evolving ecosystems. To resolve this problem, we bring back intrinsic demographic noise into the model, and show that it relaxes the conditions for phenotypic clustering and generates patterning phenomena that are distinct from deterministic Turing patterns.

\subsection{Dynamical phase transitions}

To better understand the phase transitions in the model, we consider a 1-dimensional analogue of the eco-evolutionary model in Eqs. \eqref{eq:species}-\eqref{eq:resources3}, which consists of the the following three equations:
	\begin{align}
	\dot{\rho}(x,t) &= -\nabla J(x,t) + A(\rho,r) + \sqrt{2\sigma_\rho(\rho)}\eta_\rho ~,\label{eq:1d_consumer}\\
	\dot{r}(x,t) &= B(\rho,r) + \sqrt{2\sigma_r(r)}\eta_r ~, \label{eq:1d_resource}\\
	J(x,t) &= -D(\rho)\nabla \rho + \sqrt{2\sigma_d(\rho)}\eta_j~. \label{eq:1d_diff}
\end{align}
 The mutation current $J(x,t)$ is shown in Eq. \eqref{eq:1d_diff}. The first term  of the mutation current represents the mutation rate and adaptation of consumers in trait space with a diffusion rate $D(\rho)$, and the second term is the noise.

The goal is to understand the various behaviors the complex ecosystem can exhibit depending on the ecological and evolutionary parameters. Instead of solving these coupled stochastic PDEs directly, we study the large deviations of this system. To start, we define a time- and space-integrated mutation current to be
\begin{align}
j = \frac{1}{LT}\int^T_0\int^L_0 J(x,t)~ dx dt~.
\end{align}
Using this new dynamical observable, we can construct a cumulant generating function
\begin{equation}
\label{ldt2}
\psi(\lambda)=\lim_{L,T\rightarrow\infty}\frac{1}{TL}\ln \langle e^{+\lambda TLj}\rangle~,
\end{equation}
where the brackets represent an ensemble average over the noise of the system. Drawing analogies to the free energy in equilibrium statistical physics, $LTj$ is the extensive variable with both time $T$ and the size of the phenotypic space $L$ being large. The term $\lambda$ is the conjugate intensive variable. By taking derivatives of $\psi(\lambda)$ with respect to $\lambda$, we can get all cumulants of the current $j$ \cite{touchetteLargeDeviationApproach2009,touchetteLargeDeviationApproach2013}.  

\begin{figure}[t!]
    \centering
    \includegraphics[width=0.9\linewidth]{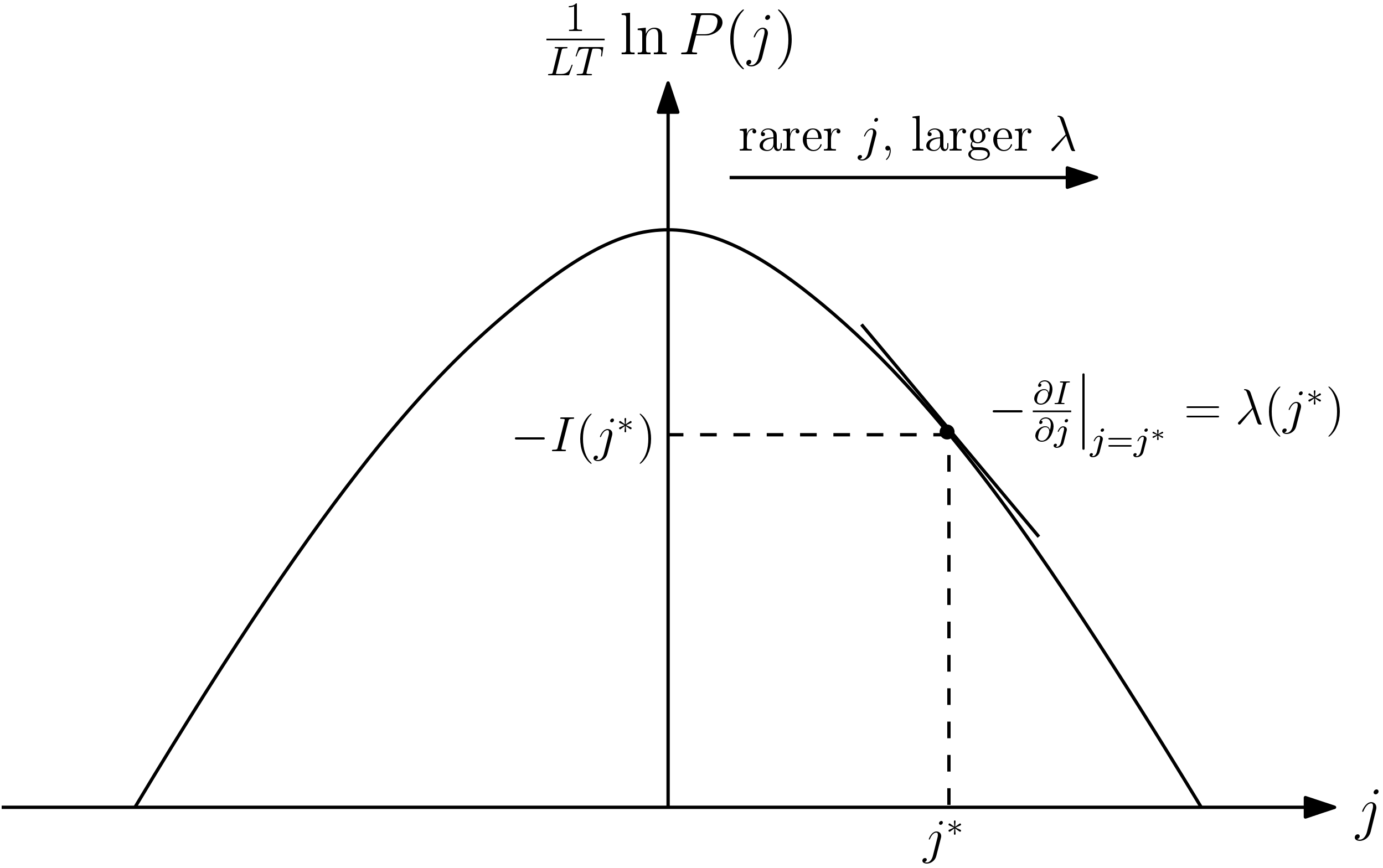}
 \label{Fi:2}
    \caption{A diagram representing the general features of the probability distribution of currents $P(j)$ and the rate function $I(j)$ along with their relation to the bias variable $\lambda$ in large deviation theory. The rate function is generally a convex (i.e. $-I(j)$ is concave) and is asymptotically related to the probability of observing a current $j$ via $\frac{1}{LT} \ln P(j) \sim - I(j) $. The rate function $I(j)$ is also the Legendre-Fenchel transform of the large deviation function $\psi(\lambda)$, where $\lambda$ is the slope of the rate function $\frac{\partial I(j)}{\partial j}$ evaluated at some current $j=j^*$. This implies that larger values of $\lambda$ correspond to more extreme and rare values of the current $j^*$, which have a lower probability of occurrence \cite{touchetteLargeDeviationApproach2009,touchetteLargeDeviationApproach2013}.}
\end{figure} 
The correspondence between $\lambda$ and $j$ can be seen by defining the probability of observing a specific $j$,
\begin{align}
    P(j) &\sim e^{-LT I(j)}~
\end{align}
where $I(j)$ is the rate function. Analogously to equilibrium statistical mechanics, one can go from the cumulant generating function to the rate function by the Legendre-Fenchel transform,
\begin{align}
\label{22}
    I(j) = \sup_\lambda \left(\lambda j - \psi(\lambda)\right)~,
\end{align}
with $\lambda^*$ being the supremum value in Eq.~\eqref{22}, and  $j(\lambda^*) = \left.\partial_\lambda \psi \right|_{\lambda = \lambda^*}$. Conversely, the intensive variable $\lambda$ is related to the rate function through the derivative $\lambda(j^*) = \left.\partial_j I\right|_{j=j^*}$. Since the rate function $I(j)$ is convex, this implies that the slope must get steeper as $j$ moves away from its average. Hence, larger values of $\lambda$ corresponds to an increasing magnitude of $j$. Using the relation between the probability $P(j)$ and $I(j)$, a larger value of $j$ is a considerably rarer occurrence and thus takes longer to observe. In summation, $\lambda$ serves to bias the original ensemble so that a larger value of $\lambda$ corresponds to observing a rarer event with a larger current $j$ \cite{touchetteLargeDeviationApproach2009,touchetteLargeDeviationApproach2013}. This relationship among $I(j), P(j)$, and $\lambda$ is demonstrated in Fig. 2.

Due to the properties of the Legendre-Fenchel transform, even if $\psi(\lambda)$ is not convex, a rate function constructed through this transformation is forced to be convex. This implies that there could be kinks in the rate functions where the derivatives of $I(j)$ start to change. These are indicative of dynamical phase transitions \cite{touchetteLargeDeviationApproach2009,touchetteLargeDeviationApproach2013}. In principle, if a phase transition is present in the tales of the distribution at a particular $j_c$, this could be associated with a critical lambda, $\lambda(j_c)$ (i.e. the dotted line in Fig.2). However, in many models, it is challenging to compute the rate function analytically. As such, we derive variational equations for the large deviation function to numerically compute dynamical phase transitions in our ecosystem model (see Eq. \eqref{varP}).

We can write Eq. \eqref{ldt2} in terms of the Martin-Siggia-Rose-Janssen-De Dominicis (MSRJD) formalism \cite{martinStatisticalDynamicsClassical1973,janssenLagrangeanClassicalField1976,dedominicisTechniquesRenormalisationTheorie1976,dedominicisFieldtheoryRenormalizationCritical1978}. Taking a saddle-point approximation, the large deviation function simplifies to an optimization of an action-like quantity $S$(see Appendix~\ref{A3}),
\begin{align}
\label{eq:CGF}
	\psi(\lambda) &=-\lim_{L,T\rightarrow\infty} \frac{1}{LT} \min_{\rho,\hat{\rho},r,\hat{r}} \int dt \int dx \ S[\rho,\hat{\rho},r,\hat{r}]~
\end{align}
where
\begin{align}
\label{eq:CGF}
S[\rho,\hat{\rho},r,\hat{r}]=\hat{\rho} \dot{\rho} +  \hat{r}\dot{r} - 	\mathcal{H}[\rho,\hat{\rho},r,\hat{r}]~,
\end{align}
and
\begin{align}
\label{action}
	\mathcal{H}[\rho,\hat{\rho},r,\hat{r}] &=  \hat{\rho} A(\rho,r) + \hat{r}B(\rho,r)- \nabla\hat{\rho}\left(D(\rho)\nabla \rho\right)\notag\\
 &\phantom{==}
 -
\lambda\left(D(\rho)\nabla \rho\right) + (\nabla \hat{\rho} + \lambda)^2 \sigma_d(\rho) \notag\\
 &\phantom{==}- \sigma_\rho(\rho)\hat{\rho}^2 - \sigma_r(r)\hat{r}^2~.
\end{align}
The action in Eq. \eqref{eq:CGF} must be minimized over the four dynamic variables, $(\rho,\hat{\rho},r,\hat{r})$,  with  $\hat{\rho}$ and $\hat{r}$ being the momentum variables conjugate to $\rho$ and $r$, respectively. The optimization problem  can be achieved by solving the Hamilton-Jacobi equations:

\begin{align}
\label{eq:rhott0}
	\partial_t \rho &= \frac{\delta}{\delta\hat{\rho}}\int dx ~\mathcal{H} \\&= A(\rho,r) + \nabla (D(\rho) \nabla \rho)- 2\nabla\sigma_d(\rho)(\nabla\hat{\rho} \label{eq:rhott}\notag \\
 &\phantom{==}- 2\nabla\sigma_d(\rho)\lambda)
	-  2\sigma_\rho(\rho)\hat{\rho}\notag\\
	\partial_{t}\hat{\rho} &= - \frac{\delta}{\delta \rho} \int dx~ \mathcal{H} \\&= -\hat{\rho} \frac{\partial A}{\partial \rho} -\hat{r} \frac{\partial B}{\partial \rho} - D(\rho) \nabla^2 \hat{\rho}\notag\\
 &\phantom{==} - \sigma_d'(\rho)(\nabla\hat{\rho} + \lambda)^2 + \sigma_\rho'(\rho)\hat{\rho}^2\notag\\
	\partial_t r &= \frac{\delta}{\delta\hat{r}} \int dx~ \mathcal{H} = B(\rho,r) - 2\sigma_r(r)\hat{r}\\
	\partial_{t}\hat{r} &= - \frac{\delta}{\delta r} \int dx~ \mathcal{H} = -\hat{\rho} \frac{\partial A}{\partial r} - \hat{r}\frac{\partial B}{\partial r} + \sigma_r'(r)\hat{r}^2~.
 \label{eq:rhott3}
\end{align}
Upon drawing a comparison between the stochastic PDEs in Eqs.~\eqref{eq:1d_consumer}-\eqref{eq:1d_diff} and the Hamilton-Jacobi equations in Eqs. \eqref{eq:rhott0}-\eqref{eq:rhott3}, we notice that the momentum variables $\hat{\rho}$ and $\hat{r}$ contain information on the noise in the consumer and resource density. 

To determine the phase diagram for our model, we expand the action around the homogeneous state $\rho_0,\hat{\rho}_0,r_0$, and $\hat{r}_0$. Since the first variation in the action must be zero, we must expand the action to second order to determine the stability of the homogeneous state. Rewriting the second order action in Fourier space with temporal and spatial frequencies $(\omega,q)$, we arrive at a $4\times4$ second order variation matrix $C_{n,m}(\omega,q)$ (see Appendix \ref{secondV2}),
\begin{equation}
\label{varP}
	\delta S = L T \sum_{n,m} V^T_{-n,-m}C_{n,m} V_{n,m}
\end{equation}
where 
\begin{align}
	V_{n,m} = \begin{pmatrix}
			\delta \rho_{n,m}\\
			\delta \hat{\rho}_{n,m}\\
			\delta r_{n,m}\\
			\delta \hat{r}_{n,m}~
		\end{pmatrix}~,
\end{align}
and
\begin{align}
 \qquad q_n = \frac{2\pi n}{L} \qquad \qquad \omega_m = \frac{2\pi m}{T}~.
\end{align}
with the $n,m$ subscript denoting the corresponding Fourier variable.

\begin{figure}[t!]
    \centering
    \includegraphics[width=0.9\linewidth]{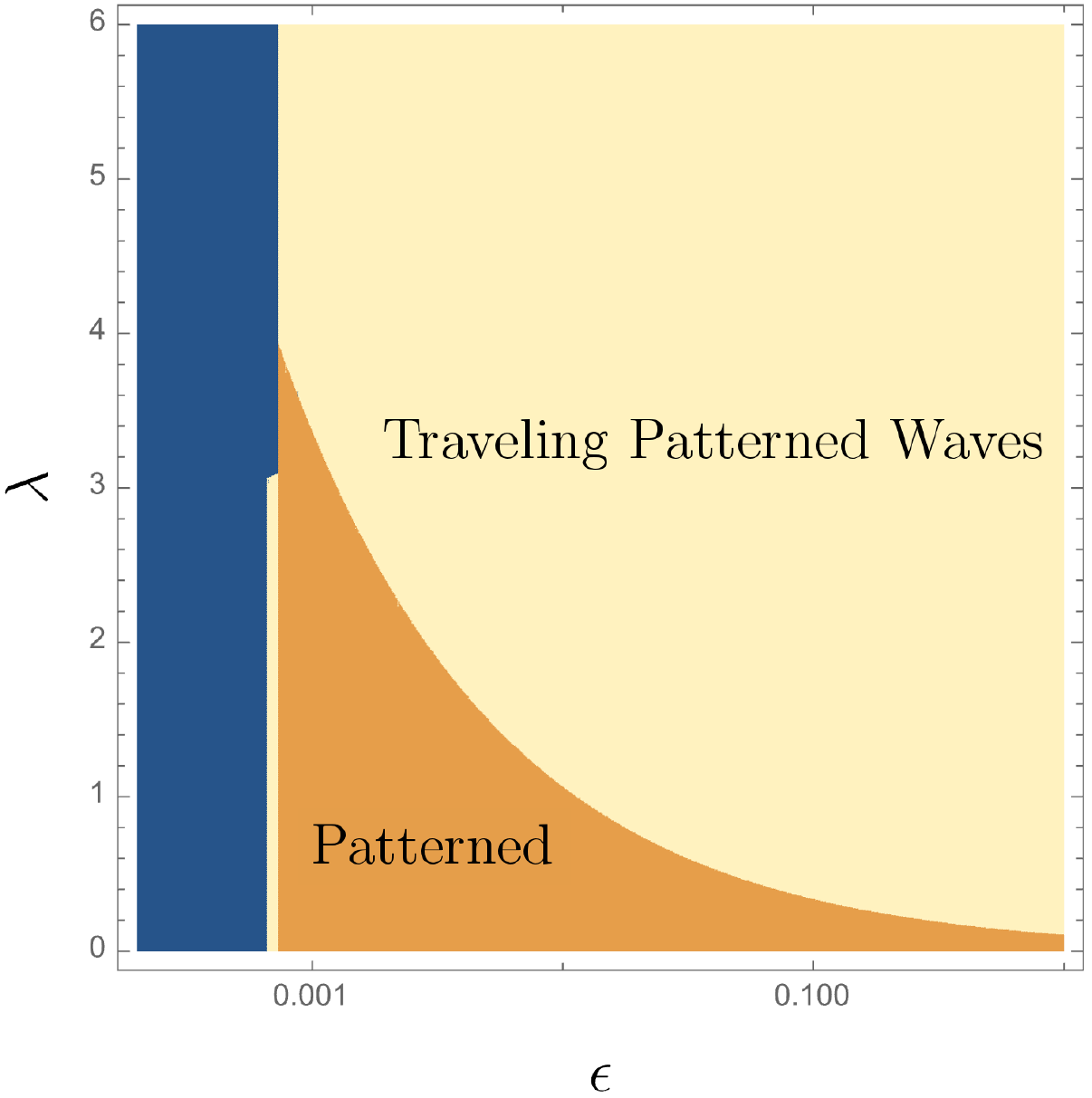}
    \caption{The Phase diagram in $\log \epsilon-\lambda$ space. The homogeneous phase (blue) occurs where the ratio of conversion to consumption of resources $\epsilon$ is very low. At low to moderate consumption rate $\epsilon$ and low biasing $\lambda$ of the current, the ecosystem exhibits a patterned phase (orange). At higher conversion rate $\epsilon$ and biasing $\lambda$, the ecosystem exhibits traveling waves (yellow). Here, the parameters are $(Z_2,D,\gamma,\kappa,\chi)= (0.9,1.0,0.1,10^5,30)$}
    \label{Fig:PhaseDiagram}
\end{figure}

\begin{figure}[t!]
    \centering
    \includegraphics[width=0.9\linewidth]{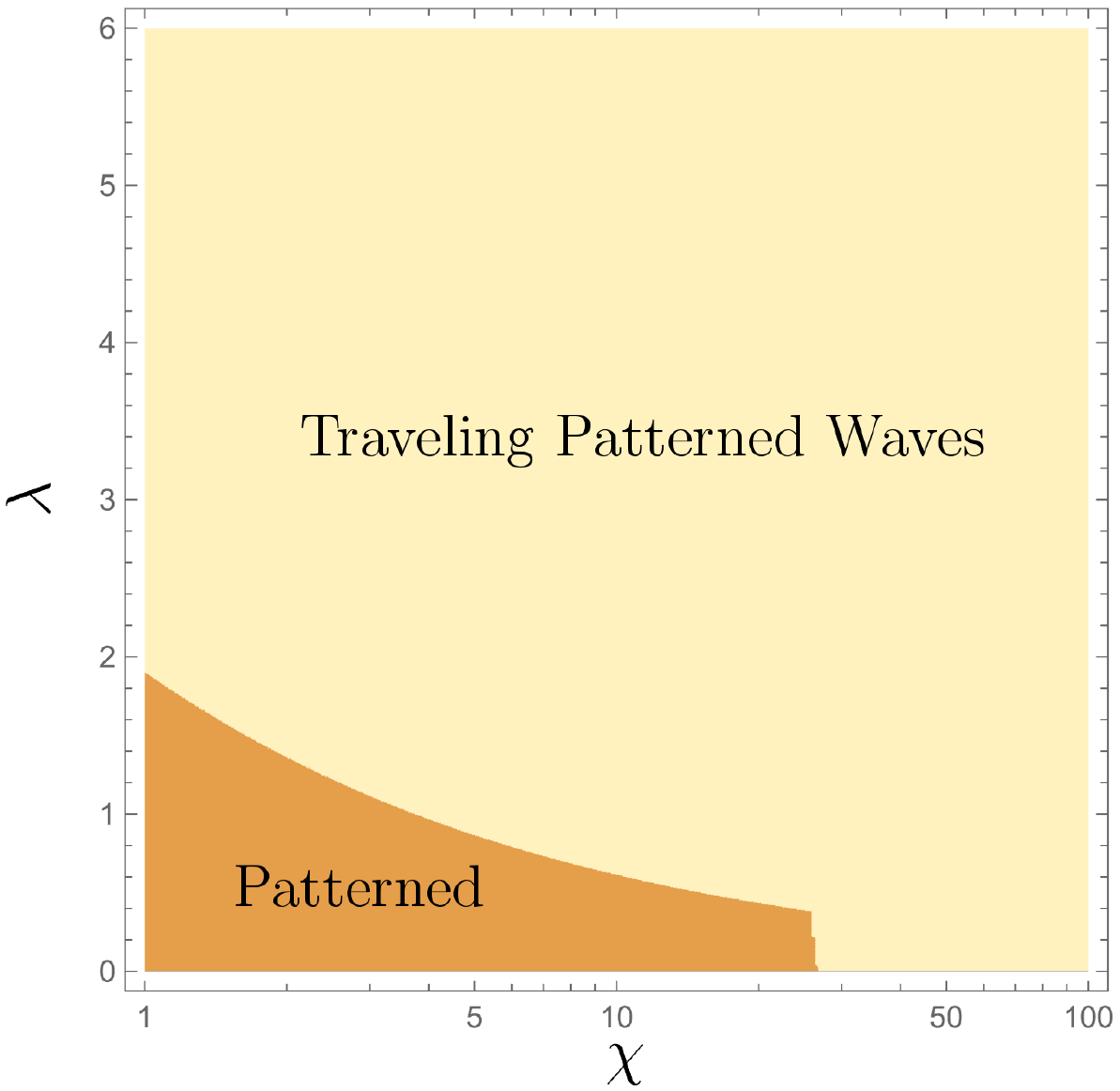}
    \caption{Phase transition in $\log \chi-\lambda$ space for a noisy consumer-resource model with mutations.  When the ratio between total consumers and total resource $\chi = \kappa/N_c$ is closer to unity, the ecosystem is more likely to be in the patterned phase (orange) and is rarely in the traveling patterned phase (yellow). On the other hand, as $\chi$ increases, the bias $\lambda$ required to observe a traveling patterned wave is lower. This indicates that the consumers are living in a relatively resource abundant environment and the consumer density profile has a high chance of being in a traveling patterned wave state. Here, the parameters are $(\epsilon,Z_2,D,\gamma,N_c) = (0.1,0.9,1.0,0.1,10^5)$. Note that the conversion efficiency is chosen to be $\epsilon=0.1$, which is an often cited ecological rule of thumb that around $10\%$ of the energy is transferred from one trophic level to the next.}
    \label{Fig:PhaseDiagram_ConsumerResourceRatio}
\end{figure}

The stability of the homogeneous state depends on the nature of the eigenvalues of the second order variation matrix $C_{n,m}(\omega,q)$~\cite{agranovMacroscopicFluctuationTheory2022}. If the $C_{n,m}(\omega,q)$ is positive definite for all frequencies $\omega$ and $q$, then any small deviations away from the steady state will decay in time, returning to the homogeneous consumer density profile. However, if the $C_{n,m}(\omega,q)$ matrix has at least one eigenvalue with a negative real part for some values of $\omega$ and $q$, then the homogeneous state is unstable and the ecosystem exhibits patterning. In the large time $T$ and system size $L$ limit, the mode corresponding to the eigenvalue with the most negative real part dictates the overall behavior of the system. If the instability is triggered by the time-independent mode $\omega =0$, then the ecosystem is patterned in niche space with little time variation~\cite{baekDynamicalSymmetryBreaking2017,baekDynamicalPhaseTransitions2018}. On the other hand, if the instability is triggered by a non-zero temporal frequency $\omega$, then the ecosystem not only clusters in phenotype space, but also co-evolves \cite{agranovMacroscopicFluctuationTheory2022, bodineauCurrentLargeDeviations2006, bodineauDistributionCurrentNonequilibrium2005,appert-rollandUniversalCumulantsCurrent2008,bertiniCurrentFluctuationsStochastic2005}. 
From this second order variation analysis, we use the determinant of $C_{n,m}(\omega,q)$ and its derivative with respect to $\omega$ to find a set of conditions under which the consumer-resource ecosystem exhibits phase transitions~\cite{agranovMacroscopicFluctuationTheory2022},
\begin{align}
\label{eqdet}
	\det(C_{n,m}(\omega,q=q_*)) &= 0 \\
 \label{eqdet2}
 \partial_\omega\det(C_{n,m}(\omega,q=q_*)) &= 0~,
\end{align}
where $q_*$ is the spatial mode that minimizes the determinant. In practice, a phase diagram is predicted by scanning over a range of parameter values and evaluating their stability according to Eqs.~\eqref{eqdet} and \eqref{eqdet2}. We find that dynamical phase transitions are sensitive to the conversion efficiency of consumed resources into biomass $\epsilon = Z_1/Z_2$, the ratio of maximum consumer to resource capacity $\chi = \kappa/N_c$, and the current bias $\lambda$. Although $\lambda$ is not a parameter or force that can be externally tuned directly, it is a fluctuation in the current $j$ and how rare it is to observe. For all cases, we consider $D=1$. In the case of free diffusion with constant diffusion rate, $\sigma_d(\rho)=D\rho$, a patterned phase is still observed and dominates the phase diagram, but phase transitions do not depend on $\lambda$. This is because dynamical phase transitions within our analysis rely on a nonlinear $\sigma_d(\rho)$ with a nonzero second derivative with respect to density. In the case of the $\sigma_d(\rho)=D\rho\left(1-\rho\right)$, we find three dynamical phases: homogeneous, patterned, and traveling patterned waves. Regarding the latter case, details of this second order variation calculation are shown in Appendix \ref{secondV2} and the corresponding phase diagrams are shown in Figs. \ref{Fig:PhaseDiagram} and \ref{Fig:PhaseDiagram_ConsumerResourceRatio}. At near-zero values of $\lambda$, the system is predominantly in its most typical state where the observed fluctuations in the current deviate slightly from the zero mean. Over time, the system continues to fluctuate and explore the tails of the distribution $P(j)$ as demonstrated in Fig. 2. If the system explores some large current fluctuations $j^*$, the corresponding $\lambda(j^*)$ may cross the phase boundary and exhibit a dynamical phase transition.

  From the phase diagram in $\log \epsilon- \lambda$ space in Fig.~\ref{Fig:PhaseDiagram}, we find that the ecosystem has a homogeneous consumer density profile only when the efficiency of conversion of consumed resources into biomass, i.e. $\epsilon=Z_1/Z_2$ is very small. The homogeneous phase is also largely independent of the current biasing $\lambda$. In this regime, very little of the resources consumed by each species is converted into biomass. Hence, the effective reproduction rate for consumers is quite low and only sufficient to maintain the population of consumers. The threshold for a subsistence level of living, $\epsilon_{th}$, is the largest value of $\epsilon$ that supports a $\lambda$-independent homogeneous phase which is set by the carrying capacity of the resources $\kappa$ and the death rate $\gamma$. In more resource rich environments with higher capacities $\kappa$, the $\epsilon_{th}$ threshold is pushed to even more minuscule values since consumers do not need to be as efficient to maintain a steady population in such an abundant environment. On the other hand, a higher death rate $\gamma$ means that the cost to maintain the internal processes of each consumer is higher. This pushes the threshold $\epsilon_{th}$ higher since more efficient usage of resources is required to escape from subsistence living and the homogeneous steady state. However, given that few species have such low efficiencies and high death rates in resource rich environments, this suggests that a homogeneous profile in niche space is quite a rarity in natural ecosystems with diverse species and abundant nutrients.

 Past this threshold efficiency $\epsilon_{th}$, the consumers are no longer living at subsistence level and can reproduce more effectively, and other phases emerge. At small efficiencies above $\epsilon_{th}$, the consumer density profile naturally forms clusters in niche space. This occurs even at low $\lambda$ biasing representing near-average current values. This runs counter to the results of a Turing analysis for the deterministic model where consumers with Gaussian interaction kernels do not spontaneously phase separate into distinct phenotypic clusters. 

 At higher values of $\lambda$, the ecosystem exhibits traveling waves where the consumers separate into distinct phenotypic clusters and co-evolve. At higher utilization of the resources, $\epsilon$, there is a decrease in the bias $\lambda$ necessary to observe the onset of traveling waves. In fact, the traveling wave regime is a predominant part of the phase diagram. This becomes even more apparent as the total resource abundance $\kappa$ increases. This suggests traveling waves with moderate speed and current are less of a rare phenomenon in natural large ecosystems when consumers are fairly efficient in converting resources into biomass. With smaller $\lambda$, an initially niche-differentiated community starts showing traveling waves after a short time. 
 
 However, there is a small discrepancy in Fig.~\ref{Fig:PhaseDiagram} at small $\epsilon$ and lower values of $\lambda$. The homogeneous phase appears to transition briefly into the traveling patterned wave phase and then into the patterned phase as $\epsilon$ increases. This may be due to numerical errors as small parameter changes can cause this region to become a part of the homogeneous phase. Furthermore, smaller values of $\epsilon$ can result in numerical instabilities and difficulties in determining whether the positive-definiteness of $C(\omega,q)$ is satisfied. 
        
From the phase diagram in $\log\chi - \lambda$ space in Fig. 4 at moderate levels of efficiency $\epsilon$, we only observe a patterned and a traveling patterned wave phase.  When the consumer and resource pool size are close to unity ($\chi\approx 1$), the consumer density profile is patterned. Consumers are competing strongly for the limited pool of resources and not all species can survive. If one species cluster begins to grow, then species with a similar phenotype are unable to outcompete the dominant species cluster and will thus decrease in abundance. As a result, the ecosystem supports distinct phenotypic clusters and niche differentiation. Away from unity at higher levels of $\chi$, the ecosystem can exhibit traveling patterned waves. As $\chi$ increases, the $\lambda$ bias necessary to observe traveling waves is lower. Hence, an ecosystem that starts off in a stationary niche differentiated phase shows signs of traveling waves after a short period of time. In fact, for values of $\chi>30$ the traveling wave pattern phase is the average behavior. This regime is the most representative of real ecosystems where a large resource pool supports a smaller, but still sizeable, consumer base.

\section{discussion}
 In this work, we developed a noisy consumer-resource model with mutations for niche space patterning. From a linear stability analysis of the deterministic version of the model, the system does not phase separate into distinct species clusters unless the interaction kernels contain negative Fourier modes. However, using large deviation theory, we show that noise in the model resolves the fine tuning problem and allows patterns to form more robustly in niche space as a rare fluctuation in the mutation current. In addition, we show that for a large portion of parameter space, an ecosystem initially in a patterned phase will eventually exhibit traveling patterned waves as a result of demographic noise. In some cases, the system will display a traveling wave phase for all $\lambda$. This is especially true in an ecosystem with a resource rich environment and sizeable pool of consumers that are moderately efficient at converting consumed resources into biomass.

Similar to the results of the generalized Lotka-Volterra model with adaptation and noise, our consumer-resource model shows that niche differentiation is a robust process that is insensitive to many details of the ecosystem \cite{rogersDemographicNoiseCan2012,rogersModesCompetitionFitness2015,rogersSpontaneousGeneticClustering2012,shnerbPatternFormationNonlocal2004}. However, our model additionally shows that the community structure is not static, but rather generically exhibits traveling patterned  waves in phenotype space. These traveling patterned waves arise as a result of the confluence of multiple factors, including biotic ones such as demographic noise and the resource utilization efficiency trait, and abiotic factors such as resource fluctuations. This suggests that many adapting ecosystems with concurrent ecological and evolutionary timescales dominated by consumer resource interactions consist of species that are constantly undergoing a co-evolutionary arms race.

A future direction for this work is to expand the large deviation function to higher order perturbations and obtain a Landau theory for the noisy consumer-resource model with mutations~\cite{baekDynamicalPhaseTransitions2018,baekDynamicalSymmetryBreaking2017}. This would be useful in determining the nature of the phase transitions by computing critical exponents. This would also provide insight into the correlations in phenotype space between various consumer species. Complementing the analytical calculation, one can also compute the large deviation function with finite size and finite time effects using numerical techniques inspired by population dynamics simulations \cite{giardinaDirectEvaluationLargeDeviation2006, giardinaSimulatingRareEvents2011,tailleurProbingRarePhysical2007,nemotoFiniteSizeScalingFirstOrder2017,nemotoPopulationdynamicsMethodMulticanonical2016}. %
Lastly, current is not the only dynamical observable that could be studied. Within the large deviations literature on lattice gas models of hard particles diffusing on a line, some have proposed entropy production and dynamical activity as useful observables to find dynamical phase transitions \cite{agranovExactFluctuatingHydrodynamics2021,agranovEntropyProductionIts2022a,agranovTricriticalBehaviorDynamical2023,jackHyperuniformityPhaseSeparation2015,dolezalLargeDeviationsOptimal2019,baekDynamicalSymmetryBreaking2017,baekDynamicalPhaseTransitions2018,grandpreEntropyProductionFluctuations2021}. For example, even in simple symmetric simple exclusion processes (SSEP), one predicts a phase transition in the dynamical activity which leads a homogeneous system to phase separate \cite{dolezalLargeDeviationsOptimal2019, derridaNonEquilibriumSteady2007,baekDynamicalPhaseTransitions2018,baekDynamicalSymmetryBreaking2017,appert-rollandUniversalCumulantsCurrent2008,brewerEfficientCharacterisationLarge2018}. Within these lattice models, optimal control forces to realize rare fluctuations have been approximated~\cite{dolezal2019large,yan2022physics,ray2018exact}. An optimal control within models like ours could shed light onto evolutionary forces that shape evolutonary dynamics~\cite{rao2022evolutionary,nourmohammad2021optimal}

Additionally, for active matter field theories with non-reciprocal interactions, one generically finds  a phase transition between a homogeneous phase and a traveling wave phase~\cite{sahaScalarActiveMixtures2020,youNonreciprocityGenericRoute2020} as well as critical exceptional points~\cite{fruchartNonreciprocalPhaseTransitions2021,el2018non,krasnok2021parity} and oscillatory transitions~\cite{cross1993pattern,cross2009pattern}.
This overlaps with our findings of a traveling wave phase in consumer-resource models, which intrinsically have non-symmetric interactions. One possible dynamical observable to investigate is the fitness flux, which was first proposed in \cite{mustonenFitnessLandscapesSeascapes2009,mustonenFitnessFluxUbiquity2010}. Due to its analogy with entropy production, it may be a fruitful avenue in identifying the possible phases and analyzing phase transitions in eco-evolutionary models.

In summary, our work shows how noise within eco-evolutionary models allow for a wide range of dynamical phase transitions to occur. These phase transitions have implications for co-evolution and niche stability. We believe our analysis opens many opportunities to study the effects of meaningful variability in more general consumer-resource models and real biological systems. 

\section{Acknowledgements}
 We thank Anne-Florence Bitbol and Stephen Martis for useful discussions. This work was supported in part by the National Institutes of Health BRAIN initiative (R01EB026943), the National Science Foundation, through the Center for the Physics of Biological Function (PHY- 1734030), and the Simons Foundation and the Sloan Foundation. JW was supported by the National Science Foundation Graduate Research Fellowship Program Grant No. DGE-2039656. DS was supported as a Simons Investigator in MMLS. T.G. was supported by the Schmidt Science Fellowship.
\newpage
\onecolumngrid

\appendix

\section{Van-Kampen Expansion}

\label{A1}
To systematically analyze the effects of intrinsic demographic noise, define an individual level model that specifies all of the different types of reactions taking place within the ecosystem. Assuming that these transitions between different population states of the ecosystem only depend on the current state of the system, then we can rely on the properties of Markov processes and capture the stochastic dynamics with a master equation. This master equation relates to the reactions shown in Fig. \ref{Fi:1} and Table I. For our set of ecological and evolutionary processes, the master equation is
	\begin{align}
		\frac{dP(\{N\},\{R\})}{dt} &= \sum_{{\bf c}} \left\{ \left(\epsilon_R^- - 1\right)\left(\frac{gR({\bf c})}{V}\right)  + \left(\epsilon_R^+ -1 \right)\left(\frac{gR({\bf c})(R({\bf c})-1)}{V^2K}\right) + \left(\epsilon_N^+-1\right) \left(\frac{dN({\bf c})}{V}\right) \right.\notag \\
		&\phantom{==}\left. + \left(\epsilon_R^+-1\right) \left(\sum_{{\bf c}'} Z_2\frac{\phi({\bf c}-{\bf c}')N({\bf c}')}{V^2} \right)R({\bf c}) + \sum_{{\bf c}'}\left( \epsilon_N^- -1\right)\left(Z_1\frac{\psi({\bf c}-{\bf c}')R({\bf c}') N({\bf c})}{V^2}\right)\right.\notag \\
		&\phantom{==} \left. + \sum_{{\bf c}'\in \mathcal{N}}\left[ \left(\epsilon_{N'}^-\epsilon_{N}^+ -1\right)\left(\frac{mN({\bf c})}{Vz}\right) + \left(\epsilon_{N'}^+\epsilon_{N}^- -1\right)\left(\frac{mN({\bf c}')}{Vz}\right) \right]  \right\} P(\{N\},\{R\})~,
	\end{align}
 where $\{N\}$ and $\{R\}$ is the set of consumer and resource abundances, respectively. Additionally, the phenotype system size is $V$, $g$ is the growth rate constant, $d$ is death rate of consumers, $z$ is the number of nearest neighbors, $K$ is the resource carrying capacity, $Z_1$ is the rate of resources to biomass conversion, $Z_2$ is the rate of consumption, and $m$ is the rate of consumer mutations. The consumption kernel $\phi({\bf c}-{\bf c}')$ represents the consumption of resource ${\bf c}$ by a consumer with a preference for ${\bf c}'$, and the conversion kernel $\psi({\bf c}-{\bf c}')$ represents the conversion of resource ${\bf c}'$ into biomass for consumers that prefer resource ${\bf c}$. Here the $\epsilon^{\pm}_N$,$\epsilon^{\pm}_R$ are step operators that increase ($+$) or decrease ($-$) the number of resource or consumer individuals by one:
	\begin{align}
		\epsilon^{\pm}_N f(\{N\},\{R\}) &= f(\ldots,N({\bf c})\pm 1,\ldots, \{ R\})\\
		\epsilon^{\pm}_R f(\{N\},\{R\}) &= f(\{ N\},\ldots,R({\bf c})\pm 1,\ldots)
	\end{align}
	Furthermore, mutations can only occur from ${\bf c}$ to another trait ${\bf c}'$ within the neighborhood $\mathcal{N}$. This can be the nearest neighbor, next-nearest neighbor, or even long-range. For simplicity, we only consider local mutations to the nearest-neighbor lattice sites.
Similar to the deterministic equations, we can non-dimensionalize the master equation as well using the same parameters:
	\begin{align}
		\frac{1}{g}\frac{dP(\{N\},\{R\})}{dt} &= \sum_{{\bf c}} \left\{ \left(\epsilon_R^- - 1\right)\left(\frac{R({\bf c})}{V}\right)  + \left(\epsilon_R^+ -1 \right)\left(\frac{R({\bf c})(R({\bf c})-1)}{V^2 K}\right) + \left(\epsilon_N^+-1\right) \left(\frac{d}{g}\right)\left(\frac{N({\bf c})}{V}\right) \right.\notag \\
		&\phantom{==}\left. + \left(\epsilon_R^+-1\right) \left(\frac{Z_2}{g}\sum_{{\bf c}'} \frac{\phi({\bf c}-{\bf c}')N({\bf c}')}{V^2} \right)R({\bf c}) \right.\notag \\
		&\phantom{==} \left.+ \left(\frac{Z_1}{g}\right)\sum_{{\bf c}'}\left( \epsilon_N^- -1\right)\left(\frac{\psi({\bf c}-{\bf c}')R({\bf c}') N({\bf c})}{V^2}\right)\right.\notag \\
		&\phantom{==} \left. + \left(\frac{mh^2}{g \sigma^2}\right) \left(\frac{\sigma^2}{h^2}\right) \sum_{{\bf c}'\in \mathcal{N}}\left[ \left(\epsilon_{N'}^-\epsilon_{N}^+ -1\right)\left(\frac{N({\bf c})}{Vz}\right) + \left(\epsilon_{N'}^+\epsilon_{N}^- -1\right)\left(\frac{N({\bf c}')}{Vz}\right) \right]  \right\}\notag \\
		&\phantom{==}\times P(\{N\},\{R\})\\
	\frac{dP(\{\tilde{N}\},\{\tilde{R}\}}{d\tau} &= \sum_{{\bf x}} \left\{ \left(\epsilon_R^- - 1\right)\left(\frac{\tilde{R}({\bf x})}{\alpha V}\right)  + \left(\epsilon_R^+ -1 \right)\left(\frac{\tilde{R}({\bf x})(\tilde{R}({\bf x})-1)}{\alpha V^2\kappa}\right) + \left(\epsilon_N^+-1\right) \gamma\left(\frac{\tilde{N}({\bf x})}{\beta V}\right) \right.\notag \\
	&\phantom{==}\left. + \left(\epsilon_R^+-1\right) \left(\sum_{{\bf x}'} \frac{\phi({\bf x}-{\bf x}')\tilde{N}({\bf x}')}{V^2\alpha} \right)\tilde{R}({\bf x}) + \sum_{{\bf x}'}\left( \epsilon_N^- -1\right)\left(\frac{\psi({\bf x}-{\bf x}')\tilde{R}({\bf x}') \tilde{N}({\bf x})}{\beta V^2}\right)\right.\notag \\
	&\phantom{==} \left. + D\left(\frac{\sigma^2}{h^2}\right) \sum_{{\bf x}'\in \mathcal{N}}\left[ \left(\epsilon_{N'}^-\epsilon_{N}^+ -1\right)\left(\frac{\tilde{N}({\bf x})}{\beta V}\right) + \left(\epsilon_{N'}^+\epsilon_{N}^- -1\right)\left(\frac{\tilde{N}({\bf x}')}{\beta V}\right) \right]  \right\}\notag \\
	&\phantom{==}\times P(\{\tilde{N}\},\{\tilde{R}\}) \label{eq:master}
	\end{align}
In summary, we have defined the following parameters:
	\begin{align*}
		\tau &= gt\\
		{\bf x} &= {\bf c}/\sigma\\
		\tilde{N} &= \beta N\\
		\tilde{R} &= \alpha R\\
		\alpha &= Z_1/g\\
		\beta &= Z_2/g\\
		\kappa &= \alpha K\\
		\gamma &= d/g\\
		D &= mh^2/g\sigma^2
	\end{align*}
	Also, note that $m$ has units of [length]$^2$[time]$^{-1}$, and it scales with the mutation rate $m$ where $h$ is the characteristic length of a phenotypic patch. The characteristic length scale $\sigma$ is the scale of interactions. 
 
 Now we need to deal with the raising and lower operators, expand them based on system size, and express them in terms of rescaled variables. 	
	Solving this master equation, however, is often difficult and approximations are necessary to gain some understanding of the ecosystem dynamics. One such approximation is the van-Kampen size expansion, which expands the master equation in terms of the system size parameter $V^{-1/2}$ and truncates at an appropriate order \cite{kampenStochasticProcessesPhysics2007}. Expanding only up to the lowest order $V^{-1/2}$ yields the mean field deterministic equations for infinite-size ecosystems. However, if we truncate at order $V^{-1}$, we convert an individual-based model of many interacting particles to a system of stochastic PDEs \eqref{eq:species2}-\eqref{eq:resources2} that can be analyzed yet still describe a large finite ecosystem. The system-size expansion begins by mapping 
	\begin{align}
		\tilde{N}({\bf x}) &= V \tilde{\rho} +\sqrt{V}\tilde{\eta} \\
		\tilde{R}({\bf x}) &= V \tilde{r} + \sqrt{V} \tilde{\xi}
	\end{align}
	and hence 
	\begin{align}
		\tilde{\eta} &= V^{-1/2}\tilde{N} - V^{1/2}\tilde{\rho}\\
		\tilde{\xi} &= V^{-1/2}\tilde{R} - V^{1/2}\tilde{r}
	\end{align}
	To rewrite the probability distribution $P(\{\tilde{N}\},\{\tilde{R}\})$ in terms of the variables $\tilde{\eta},\tilde{\xi}$, we define
	\begin{align}
	 \Pi\left(\{\tilde{\eta}\},\{\tilde{\xi}\}\right) =VP(\{\tilde{N}\},\{\tilde{R}\})~.
	\end{align}
Substituting this back into $dP/dt$, we have 
	\begin{align}
		\frac{\partial P}{\partial \tau} &= \frac{1}{V}\left(\frac{\partial \Pi}{\partial \tau} + \sum_{{\bf x}}\left( \frac{\partial \Pi}{\partial \tilde{\eta}}\frac{\partial \tilde{\eta}}{\partial \tau} +\frac{\partial \Pi}{\partial \tilde{\xi}}\frac{\partial \tilde{\xi}}{\partial \tau} \right)\right)\\
			&= \frac{1}{V}\left(\frac{\partial \Pi}{\partial \tau} - \sqrt{V}\sum_{{\bf x}} \left(\frac{\partial P}{\partial \tilde{\eta}} (\partial_t \tilde{\rho})-\frac{\partial \Pi}{\partial \tilde{\xi}}(\partial_t \tilde{r})\right)\right)\\
			&= \frac{1}{V}\frac{\partial \Pi}{\partial \tau} - \frac{1}{\sqrt{V}}\sum_{{\bf x}} \left(\frac{\partial \Pi}{\partial \tilde{\eta}} (\partial_t \tilde{\rho})+\frac{\partial \Pi}{\partial \tilde{\xi}}(\partial_t \tilde{r})\right)
	\end{align}

	Now, we turn our attention to the right-hand side of the master equation and expand the step operators in terms of the parameter $V^{-1/2}$. Recall that the action of a step operator $\epsilon^{\pm \ell}$ on a function $f(N)$ is 
	\begin{align}
		\epsilon^{\pm \ell} f(N) = f(N\pm \ell) = \sum_{k=0}^\infty \frac{\ell^k}{k!}\frac{d^kf}{dN^k} = \sum^\infty_{k=0} \frac{1}{k!} \left[\ell \frac{\partial}{\partial N}\right]^k f(N) = e^{\ell \partial_N} f(N)
	\end{align}
	where $\ell$ is small. Since $N = V\rho + \sqrt{V}\eta$, $\tilde{N} = V\tilde{\rho} + \sqrt{V}\tilde{\eta}$, and $\tilde{N}= bN$, then we find that $\partial_{N} = (a/\sqrt{V})\partial_{\tilde{\eta}}$ and we have 
	\begin{align}
			\epsilon^{\pm \ell}-1 =\pm \frac{1}{\sqrt{V}}\frac{\partial}{\partial\eta} + \frac{1}{V}\frac{\partial}{\partial \eta^2} =   \pm \frac{b}{\sqrt{V}}\frac{\partial}{\partial\tilde{\eta}} + \frac{b^2}{V}\frac{\partial}{\partial \tilde{\eta}^2}
	\end{align}
	Notice that this is the same as saying that if $N$ changes by 1, then $\tilde{N}$ changes by $b$. By a similar logic, we can derive an equivalent expression for the operators acting on functions of the resource population. In summary, we have
	\begin{align}
		\epsilon^{\pm}_{N} - 1 &= \pm \frac{\beta}{\sqrt{V}}\frac{\partial}{\partial \tilde{\eta}} + \frac{\beta^2}{2V}\frac{\partial^2}{\partial \tilde{\eta}^2} \label{eq:Nchange}\\
		\epsilon^{\pm}_{R} - 1 &= \pm \frac{\alpha}{\sqrt{V}} \frac{\partial}{\partial \tilde{\xi}} + \frac{\alpha^2}{2V}\frac{\partial^2}{\partial \tilde{\xi}^2}\label{eq:Rchange}\\
		\epsilon_{N'}^-\epsilon_{N}^+ -1 &=  \frac{\beta}{\sqrt{V}} \left(\frac{\partial}{\partial \tilde{\eta}}- \frac{\partial}{\partial \tilde{\eta}'}\right) + \frac{\beta^2}{2V}\left(\frac{\partial}{\partial \tilde{\eta}}- \frac{\partial}{\partial \tilde{\eta}'}\right)^2 \label{eq:NNchange1} \\
		\epsilon_{N'}^+\epsilon_{N}^- -1  &= \frac{\beta}{\sqrt{V}} \left(\frac{\partial}{\partial \tilde{\eta}'}- \frac{\partial}{\partial \tilde{\eta}}\right) + \frac{\beta^2}{2V}\left(\frac{\partial}{\partial \tilde{\eta}'}- \frac{\partial}{\partial \tilde{\eta}}\right)^2 \label{eq:NNchange2}
	\end{align}
	Notice that there is a factor of $\alpha^{-1}$ multiplying the resource number $\tilde{R}$ and $\beta^{-1}$ multiplying the consumer population $\tilde{N}$ in the master equation in \eqref{eq:master}. If we distribute this factor of $\alpha$ or $\beta$ and absorb it into the step operators in equations \eqref{eq:Nchange}-\eqref{eq:NNchange2}, we have 
	\begin{align}
		\frac{1}{\beta}\left(\epsilon^{\pm}_{N} - 1\right) &= \pm \frac{1}{\sqrt{V}}\frac{\partial}{\partial \tilde{\eta}} + \frac{\beta}{2V}\frac{\partial^2}{\partial \tilde{\eta}^2}\\
		\frac{1}{\alpha}\left(\epsilon^{\pm}_{R} - 1\right) &=  \pm \frac{1}{\sqrt{V}} \frac{\partial}{\partial \tilde{\xi}} + \frac{\alpha}{2V}\frac{\partial^2}{\partial \tilde{\xi}^2}\\
		\frac{1}{\beta}\left(\epsilon_{N'}^-\epsilon_{N}^+ -1 \right)&=  \frac{1}{\sqrt{V}} \left(\frac{\partial}{\partial \tilde{\eta}}- \frac{\partial}{\partial \tilde{\eta}'}\right) + \frac{\beta}{2V}\left(\frac{\partial}{\partial \tilde{\eta}}- \frac{\partial}{\partial \tilde{\eta}'}\right)^2  \\
		\frac{1}{\beta}\left(\epsilon_{N'}^+\epsilon_{N}^- -1 \right) &=  \frac{1}{\sqrt{V}} \left(\frac{\partial}{\partial \tilde{\eta}'}- \frac{\partial}{\partial \tilde{\eta}}\right) + \frac{\beta}{2V}\left(\frac{\partial}{\partial \tilde{\eta}'}- \frac{\partial}{\partial \tilde{\eta}}\right)^2 
	\end{align}
	
	Using these expansions, we now list the various contributing terms to the master equation and group terms at order $V^{-1/2}$ and $V^{-1}$:
	\begin{enumerate}
		\item Growth of resources:
			\begin{align*}
			\frac{1}{\alpha}(\epsilon_R^- - 1)\left(\frac{\tilde{R}({\bf x})}{V}\right) =	- \frac{1}{\sqrt{V}} \left(\tilde{r} \frac{\partial}{\partial \tilde{\xi}}\right)+ \frac{1}{V}\left(- \frac{\partial}{\partial \tilde{\xi}}\tilde{\xi} + \frac{a\tilde{r}}{2} \frac{\partial^2}{\partial\tilde{\xi}^2} \right)
			\end{align*}
			
		\item Decay of resources:
			\begin{align*}
				\frac{1}{\alpha}(\epsilon_R^+ - 1)\left(\frac{\tilde{R}({\bf x})(\tilde{R}({\bf x})-1)}{V^2\kappa}\right) = -\frac{1}{\sqrt{V}}\left(-\frac{\tilde{r}^2}{\kappa}\frac{\partial}{\partial \tilde{\xi}}\right) + \frac{1}{V}\left(\frac{2\tilde{r}}{\kappa}\frac{\partial}{\partial \tilde{\xi}}\tilde{\xi} + \frac{\alpha\tilde{r}^2}{2\kappa}\frac{\partial^2}{\partial\tilde{\xi}^2} \right)
			\end{align*}
			
		\item Death of consumer:
			\begin{align*}
				\frac{1}{\beta}(\epsilon_N^+-1)\left(\frac{\gamma \tilde{N}({\bf x})}{V}\right) =  -\frac{1}{\sqrt{V}}\left(-\gamma \tilde{\rho} \frac{\partial}{\partial \tilde{\eta}} \right) + \frac{1}{V}\left(\gamma \frac{\partial}{\partial \tilde{\eta}}\tilde{\eta} + \frac{\gamma \beta \tilde{\rho}}{2}\frac{\partial^2}{\partial\tilde{\eta}^2}\right)
			\end{align*}
			
		\item Consumption of resource ${\bf x}$ by consumer of trait ${\bf x}'$
			\begin{align*}
				\frac{1}{\alpha}(\epsilon_{R}^+-1)\left( \frac{\phi({\bf x}-{\bf x}')\tilde{N}({\bf x}')\tilde{R}({\bf x})}{V^2}\right) &= -\frac{1}{\sqrt{V}}\left(-\phi({\bf x}-{\bf x}')\tilde{\rho}({\bf x}') \tilde{r}({\bf x}) \frac{\partial}{\partial\tilde{\xi}} \right)\\
				&\phantom{=} + \frac{1}{V}\left[\phi({\bf x}-{\bf x}')\frac{\partial}{\partial \tilde{\xi}} \left(\tilde{\eta}({\bf x}')\tilde{r}({\bf x})  + \tilde{\rho}({\bf x}')\tilde{\xi}({\bf x})\right) \right.\\
					&\phantom{=} \left. + \frac{\alpha}{2} \phi({\bf x}-{\bf x}') \tilde{\rho}({\bf x}')\tilde{r}({\bf x})\frac{\partial^2}{\partial \tilde{\xi}^2}\right]
			\end{align*}
		
		\item Growth of consumer of trait ${\bf x}$ by consuming resource ${\bf x}'$:
			\begin{align*}
				\frac{1}{\beta}(\epsilon_{N}^- - 1) \left(\frac{\psi({\bf x}-{\bf x}')\tilde{R}({\bf x}')\tilde{N}({\bf x})}{V^2} \right)&= -\frac{1}{\sqrt{V}}\left(\psi({\bf x}-{\bf x}')\tilde{r}({\bf x}') \tilde{\rho}({\bf x}) \frac{\partial}{\partial \tilde{\eta}} \right) \\
				&\phantom{=}  + \frac{1}{V}\left[-\psi({\bf x}-{\bf x}')  \frac{\partial }{\partial \tilde{\eta}}\Big(\tilde{\rho}({\bf x})\tilde{\xi}({\bf x}') + \tilde{\eta}({\bf x})\tilde{r}({\bf x}')\Big)\right.\\
				&\phantom{=} + \left. \frac{\beta}{2}\psi({\bf x}-{\bf x}')\tilde{r}({\bf x}')\tilde{\rho}({\bf x}) \frac{\partial^2}{\partial \tilde{\eta}^2} \right]
			\end{align*} 
			
		\item Consumer mutations:
			\begin{align*}
				&\phantom{=} \left(\frac{\sigma^2}{h^2}\right)\sum_{{\bf x}}\sum_{{\bf x}'\in \mathcal{N}} \left[(\epsilon_{N'}^-\epsilon_N^+ - 1) \left(\frac{D\tilde{N}({\bf x})}{zV\beta}\right) + (\epsilon_{N'}^+ \epsilon_N^- - 1)\left(\frac{D\tilde{N}({\bf x}')}{zV\beta }\right)\right]\\
				&= \left(\frac{\sigma^2}{h^2}\right)\sum_{{\bf x}}\left\{\frac{1}{\sqrt{V}}\left(\frac{D}{z}\sum_{{\bf x}' \in \mathcal{N}}\left( \tilde{\rho}({\bf x}')- \tilde{\rho}({\bf x})\right)\left(\frac{\partial}{\partial \tilde{\eta}'} - \frac{\partial}{\partial \tilde{\eta}}\right) \right) + \frac{1}{V}\left[\left(\frac{D}{z} \sum_{{\bf x}'\in \mathcal{N}}\left(\frac{\partial}{\partial \tilde{\eta}'} - \frac{\partial}{\partial \tilde{\eta}}\right) (\tilde{\eta}({\bf x}')- \tilde{\eta}({\bf x})) \right)\right.\right.\\
				&\phantom{=} \left.\left.+ \left(\frac{D\beta}{2z}\sum_{{\bf x}'\in \mathcal{N}}\left( \tilde{\rho}({\bf x}') + \tilde{\rho}({\bf x})\right)\left(\frac{\partial}{\partial\tilde{\eta}'}- \frac{\partial}{\partial\tilde{\eta}} \right)^2\right) \right]\right\}\\
				&=\left( \frac{\sigma^2}{h^2}\right)\sum_{{\bf x}}\left\{-\frac{1}{\sqrt{V}} \left(D\left(\frac{2}{z}\sum_{{\bf x}'\in \mathcal{N}}(\tilde{\rho}({\bf x}')-\tilde{\rho}({\bf x}))\right)\frac{\partial}{\partial \tilde{\eta}} \right) + \frac{1}{V}\left[- D\frac{\partial}{\partial\tilde{\eta}}\left(\frac{2}{z}\sum_{{\bf x}' \in \mathcal{N} } (\tilde{\eta}({\bf x}') - \tilde{\eta}({\bf x}))\right) \right. \right.\\
				&\phantom{=} + \left.\left. \frac{D\beta}{z}\left(\sum_{{\bf x}'}\left(\tilde{\rho}({\bf x}') + \tilde{\rho}({\bf x})\right)\right)  \frac{\partial^2}{\partial \tilde{\eta}^2} - \frac{D\beta}{z} \sum_{{\bf x}'}\left( (\tilde{\rho}({\bf x}') + \tilde{\rho}({\bf x})) \frac{\partial^2}{\partial \tilde{\eta} \partial \tilde{\eta}'} \right)\right]\right\}\\
				&= \sum_{{\bf x}} \left\{ - \frac{1}{\sqrt{V}}\left( D \nabla^2 \tilde{\rho}({\bf x}) \frac{\partial}{\partial \tilde{\eta}}\right) + \frac{1}{V}\left( -D\frac{\partial}{\partial \tilde{\eta}}\left(\nabla^2 \tilde{\eta}({\bf x})\right) +\frac{D\beta}{z}\left(\frac{\sigma^2}{h^2}\right)\left(\sum_{{\bf x}'}\left(\tilde{\rho}({\bf x}) + \tilde{\rho}({\bf x}')\right)\right) \frac{\partial^2}{\partial \tilde{\eta}^2} \right.\right.\\
				&\phantom{=}  \left.\left.  - \frac{D\beta}{z}\left(\frac{\sigma^2}{h^2}\right) \sum_{{\bf x}'} \left((\tilde{\rho}({\bf x}') + \tilde{\rho}({\bf x})) \frac{\partial^2}{\partial \tilde{\eta} \partial \tilde{\eta}'} \right)  \right) \right\}
			\end{align*}
			
	\end{enumerate}
	For the first term in the mutation expression, there is an extra factor of two because we obtain the same expression $(\tilde{\rho}({\bf x}') - \tilde{\rho}({\bf x}))\partial_{\tilde{\eta}}$ from doing the same calculations on all neighbors of ${\bf c}$.  The same logic applies for the second and third terms of (6).
	
	Now, identifying all terms of order $V^{-1/2}$ on the left-hand-side and right-hand-side of the master equation, we derive the deterministic MacArthur consumer-resource equations with mutations and nonlocal trait space interactions:
	\begin{align}
		\frac{\partial \tilde{\rho}({\bf x})}{\partial \tau} &= \tilde{\rho}({\bf x})\left(\sum_{{\bf c}'} \psi({\bf x}'-{\bf x}) \tilde{r}({\bf x}') - \gamma\right) + D \nabla^2 \tilde{\rho}({\bf x})\\
		\frac{\partial \tilde{r}({\bf x})}{\partial \tau} &= \tilde{r}({\bf x}) \left( 1- \frac{\tilde{r}({\bf x})}{\kappa} - \sum_{{\bf x}'} \left(\phi({\bf x}-{\bf x}')\right) \tilde{\rho}({\bf x}')\right)
	\end{align}

\section{Non-dimensionalization}
\label{A2}
To better understand the overall dynamics of these stochastic partial differential equations, it is advantageous to rescaling the equations by rewriting them in terms of effective variables. To start, we redefine our model with units to be the following:
	\begin{align*}
		\frac{d\rho({\bf c})}{dt} &= \rho({\bf c}) \left[\frac{Z_1}{g} \int \psi({\bf c}-{\bf c}') r({\bf c}') d{\bf c}' - \frac{\gamma}{g}\right] + \frac{D}{g}\nabla^2_{\bf c}\rho({\bf c}) + \eta(t) \\
		\frac{dr({\bf c})}{dt} &= r({\bf c})\left[ 1 - \frac{1}{\kappa}r({\bf c})- \frac{Z_2}{g}\int \phi({\bf c}-{\bf c}') \rho({\bf c}') d{\bf c}'  \right] + \xi(t)~.
	\end{align*}
Then, we define a unitless time $t \rightarrow gt$, scaling the time with the growth rate of the resources. If $\sigma$ is the trait space scale, then we should set ${\bf x} = {\bf c}/\sigma$. This allows us to rewrite the above as 
	\begin{align*}
		\frac{d\rho({\bf x})}{dt} &= \rho({\bf x}) \left[\frac{Z_1 }{g} \int \psi({\bf x}-{\bf x}') r({\bf x}') d{\bf x}' - \frac{\gamma}{g}\right] + \frac{D}{g\sigma^2} \nabla^2_{\bf x}\rho({\bf x}) + \eta(t) \\
		\frac{dr({\bf x})}{dt} &= r({\bf x})\left[ 1 - \frac{1}{\kappa}r({\bf x})- \frac{Z_2 }{g}\int \phi({\bf x}-{\bf x}') \rho({\bf x}') d{\bf x}'  \right] + \xi(t) 
	\end{align*}
	with $\psi({\bf x}) = \sigma^M \psi({\bf c}({\bf x}))$ and $\phi({\bf x}) = \sigma^M \phi({\bf c}({\bf x}))$ to ensure that $\psi$ and $\phi$ are still normalized distributions. Re-defining parameters $Z_1 \rightarrow Z_1/g$, $Z_2\rightarrow Z_2/g$, $\gamma \rightarrow \gamma/g$ and $D = D/g\sigma^2$, we have
	\begin{align}
		\frac{d\rho({\bf x})}{dt} &= \rho({\bf x}) \left[Z_1\int \psi({\bf x}-{\bf x}') r({\bf x}') d{\bf x}' - \gamma \right] + D \nabla^2_{\bf x}\rho({\bf x}) + \eta({\bf x},t) \label{eq:species2} \\
		\frac{dr({\bf x})}{dt} &= r({\bf x})\left[ 1 - \frac{1}{\kappa}r({\bf x})- Z_2\int \phi({\bf x}-{\bf x}') \rho({\bf x}') d{\bf x}'  \right] + \xi({\bf x},t)  \label{eq:resources2}
	\end{align}
     where the multiplicative noise terms are given by 
     \begin{align}
         \eta({\bf x},t) &= \sqrt{2\sigma_\rho(\rho)}\eta_\rho(t) + \nabla\left(\sqrt{2\sigma_d(\rho)}\eta_j(t) \right) \notag \\
         &= \sqrt{\frac{1}{N}\rho({\bf x}) \left[Z_1\int \psi({\bf x}-{\bf x}') r({\bf x}') d{\bf x}' + \gamma \right]} \eta_\rho(t) + \nabla\left(\sqrt{2 \sigma_d(\rho) }\eta_j(t)\right)\\
        \xi({\bf x,t}) &= \sqrt{2\sigma_r(r)}\eta_r(t) \notag\\
        &=\sqrt{\frac{1}{\kappa} r({\bf x})\left[ 1 + \frac{1}{\kappa}r({\bf x}) +Z_2 \int \phi({\bf x}-{\bf x}') \rho({\bf x}') d{\bf x}'  \right]} \eta_r(t)
     \end{align}
     Here, $\sigma_d(\rho)$ is the noise from the diffusion of consumers in niche space, which arises from consumer mutations and adaptations. As stated in the main text, this noise can take various forms depending on the model assumptions and the interactions in the system. Here, we consider the forms $D(\rho) \rho$ and $D\rho(1-\rho)$. The former arises for freely diffusing consumers in niche space with a diffusion rate of $D(\rho)$. The latter can arise from competitive exclusion of new mutants that have similar traits to a highly abundant strong competitor or from density dependent mutation rates. Furthermore, $N$ is the total population size of the consumers. If it is fixed, it represents the carrying capacity of the community. Furthermore, the noise terms follow the usual Wiener process:
     \begin{align}
         \langle \eta_\rho\rangle = \langle \eta_j\rangle = \langle \eta_r\rangle &= 0\\
         \langle \eta_\rho({\bf x},t)\eta_\rho({\bf x}',t')\rangle = \langle \eta_j({\bf x},t)\eta_j({\bf x},t')\rangle = \langle \eta_r({\bf x},t) \eta_r({\bf x}',t')\rangle &= \delta({\bf x}-{\bf x}')\delta(t-t')
     \end{align}
        
\section{Martin-Siggia-Rose Formulation}
\label{A3}

Instead of trying to tilt the ensemble and write the biased Fokker-Planck equation directly, we can use the MSRJD formalism to derive the correct equation. Furthermore, since this method relies on path integrals, we can easily compute desired statistics without deriving the stationary distribution (which is usually very difficult to obtain). We have equations of the form 
\begin{align}
	\dot{\rho}(x,t) &= -\nabla J(x,t) + A(\rho,r) + \sqrt{2\sigma_\rho(\rho)}\eta_\rho \\
	\dot{r}(x,t) &= B(\rho,r) + \sqrt{2\sigma_r(r)}\eta_r\\
	J(x,t) &= -D(\rho) \nabla \rho + \sqrt{2\sigma_d(\rho)}\eta_j
\end{align}
Here we use $A(\rho,r)$ and $B(\rho,r)$ to capture the birth-death dynamics of consumers and resources:
\begin{align}
    A(\rho,r) &= \rho(x) \left(Z_1 \int \psi(x-x') r(x') dx' - \gamma\right) \\
    B(\rho,r) &= r(x) \left(1- \frac{r(x)}{\kappa} - Z_2 \int \phi(x-x') \rho(x') dx'\right)
\end{align}

The path probability of this system is given by
\begin{align}
	P &= \int D[\eta_\rho,\eta_r,\eta_j,\rho,r,j]~ \delta\left(\dot{\rho} - A(\rho,r) - \sqrt{2\sigma_\rho(\rho)}\eta_\rho \right)\delta\left(j + D(\rho)\nabla \rho - \sqrt{2\sigma_d(\rho)}\eta_j\right)\notag\\
	&\phantom{===} \delta\left(\dot{r}- B(\rho,r)-\sqrt{2\sigma_r(r)}\right) \exp\left(-LT\int\int dxdt ~ \left(\frac{\eta_j^2}{2} + \frac{\eta_\rho^2}{2} + \frac{\eta_r^2}{2}\right)\right)
\end{align}
To proceed, we will use the definition of a delta function to rewrite
\begin{align}
	&\delta\left(\dot{\rho}+\nabla j- A(\rho,r) - \sqrt{2\sigma_\rho(\rho)}\eta_\rho\right) = \notag\\
	&\phantom{==}\int D[\hat{\rho}]~\exp\left(-LT\int\int dx dt ~ \hat{\rho}\left(\dot{\rho}+\nabla j- A(\rho,r) - \sqrt{2\sigma_\rho(\rho)}\eta_\rho\right) \right)
\end{align}
where $\hat{\rho}$ is integrated along the imaginary axis. We employ the same method for the delta function $\delta\left(\dot{r}- B(\rho,r)-\sqrt{2\sigma_r(r)}\right)$, which allows us the express the path probability as 
\begin{align}
	  P &= \int D[\eta_\rho,\eta_r,\eta_j,\rho, \hat{\rho},r,\hat{r},j]~\delta\left(j + D(\rho)\nabla \rho - \sqrt{2\sigma_d(\rho)}\eta_j\right)~\notag\\
	&\phantom{===} \exp\left[-LT\int\int dx dt ~ \hat{\rho}\left(\dot{\rho}+\nabla j- A(\rho,r) - \sqrt{2\sigma_\rho(\rho)}\eta_\rho\right) \right.\notag\\
	&\phantom{====}\left. \phantom{\int}+ \hat{r}\left(\dot{r} - B(\rho,r) - \sqrt{2\sigma_r(r)}\eta_r\right) + \left(\frac{\eta_j^2}{2} + \frac{\eta_\rho^2}{2} + \frac{\eta_r^2}{2}\right)\right]
\end{align}
To compute statistics on the consumer's current in the system, we must compute the generating functional $z(\lambda) = \langle e^{LT\lambda J}\rangle$ where $\lambda$ is the biasing field and $J$ is the time averaged current
\begin{align}
	J = \frac{1}{LT}\int^T_0\int^L_0 j(x,t)~ dx dt
\end{align}

In order to bias for current, we must compute the generating functional $\psi(\lambda) = \langle e^{LT\lambda J}\rangle$. Note that $\lambda$ is the conjugate variable to the current. Using this generating functional approach, we can compute all statistics for both dynamical quantities by taking derivatives of $z(\lambda)$. 

Using the expression for the path probability and integrating over $j$, we have
\begin{align}
 z(\lambda) &= \int D[\eta_\rho,\eta_r,\eta_j,\rho, \hat{\rho},r,\hat{r},j]~\delta\left(j + D(\rho)\nabla \rho - \sqrt{2\sigma_d(\rho)}\eta_j\right)~\notag\\
 &\phantom{===} \exp\left[-LT\int\int dx dt ~ \hat{\rho}\left(\dot{\rho}+\nabla j- A(\rho,r) - \sqrt{2\sigma_\rho(\rho)}\eta_\rho\right) \right.\notag\\
 &\phantom{==}\left. \phantom{\int}+ \hat{r}\left(\dot{r} - B(\rho,r) - \sqrt{2\sigma_r(r)}\eta_r\right) + \left(\frac{\eta_j^2}{2} + \frac{\eta_\rho^2}{2} + \frac{\eta_r^2}{2}\right) - \lambda j \right]\\
 &= \int D[\eta_\rho,\eta_r,\eta_j,\rho, \hat{\rho},r,\hat{r}]~ \exp\left[-LT\int\int dx dt ~ \hat{\rho}\left(\dot{\rho}- A(\rho,r) - \sqrt{2\sigma_\rho(\rho)}\eta_\rho\right)   \right.\notag\\
 &\phantom{==} \left.\phantom{\int}+ \hat{r}\left(\dot{r} - B(\rho,r) - \sqrt{2\sigma_r(r)}\eta_r\right) + \left(\frac{\eta_j^2}{2} + \frac{\eta_\rho^2}{2} + \frac{\eta_r^2}{2}\right) \right.\notag\\
 &\phantom{==} \phantom{\int}- (\hat{\rho} \nabla - \lambda) \left(D(\rho)\nabla \rho -  \sqrt{2\sigma_d(\rho)}\eta_j\right)\Bigg]\\
 &=  \int D[\eta_\rho,\eta_r,\eta_j,\rho, \hat{\rho},r,\hat{r}]~ \exp\left[-LT\int\int dx dt ~ \hat{\rho}\left(\dot{\rho}- A(\rho,r) - \sqrt{2\sigma_\rho(\rho)}\eta_\rho\right)  \right.\notag\\
 &\phantom{==} \left.\phantom{\int}+\hat{r}\left(\dot{r} - B(\rho,r) - \sqrt{2\sigma_r(r)}\eta_r\right) + \left(\frac{\eta_j^2}{2} + \frac{\eta_\rho^2}{2} + \frac{\eta_r^2}{2}\right) \right.\notag\\
 &\phantom{==} \phantom{\int}+ ( \nabla\hat{\rho} + \lambda) \left(D(\rho)\nabla \rho -  \sqrt{2\sigma_d(\rho)}\eta_j\right)\Bigg]
\end{align}
In the final line, we used integration by parts on the last term and the fact that $\hat{\rho}$ is zero at long times. This allows us the integrate out all of the noise terms $\eta_j, \eta_\rho,$ and $\eta_r$, yielding
\begin{align}
	z(\lambda) &=  \int D[\rho, \hat{\rho},r,\hat{r}]~ \exp\left[-LT\int\int dx dt ~ \hat{\rho}\left(\dot{\rho}- A(\rho,r) \right) + \hat{r}\left(\dot{r} - B(\rho,r)\right)\right.\notag\\
	&\phantom{==} \left. \phantom{\int}+ ( \nabla\hat{\rho} + \lambda) \left(D(\rho)\nabla \rho\right)   - (\nabla \hat{\rho} + \lambda)^2 \sigma_d(\rho) + \sigma_\rho(\rho)\hat{\rho}^2 + \sigma_r(r)\hat{r}^2 \right]
\end{align}
Hence, the classical action for this system is 
\begin{align}
	\mathcal{S}[\rho,\hat{\rho},r,\hat{r}] = \int\int dx dt ~ \hat{\rho} \dot{\rho} +  \hat{r}\dot{r} - 	\mathcal{H}[\rho,\hat{\rho},r,\hat{r}]
\end{align}
with the path probability given by
\begin{align}
	-\ln P \sim LT\mathcal{S}
\end{align}
and Hamiltonian
\begin{align}
	\mathcal{H}[\rho,\hat{\rho},r,\hat{r}] &=  \hat{\rho} A(\rho,r) + \hat{r}B(\rho,r) - (\nabla\hat{\rho} + \lambda) \left(D(\rho)\nabla \rho\right)  \notag\\
 &\phantom{==} + (\nabla \hat{\rho} + \lambda)^2 \sigma_d(\rho) +\sigma_\rho(\rho)\hat{\rho}^2+ \sigma_r(r)\hat{r}^2 
\end{align}

The large deviation function is 
\begin{align}
	\psi(\lambda) = -\lim_{L,T\rightarrow\infty} \frac{1}{LT} \min_{\rho,\hat{\rho},r,\hat{r}} \int^T_0 dt \int^L_0 dx ~ \hat{\rho} \dot{\rho} +  \hat{r}\dot{r} - 	\mathcal{H}[\rho,\hat{\rho},r,\hat{r}]
\end{align}
where we have to minimize over $\rho,\hat{\rho},r,\hat{r}$. This can be accomplished by using a saddlepoint apporximation, solving the Hamilton-Jacobi equations, and substituting the results. The Hamilton-Jacobi equations are
\begin{align}
	\partial_t \rho &= \frac{\delta}{\delta \hat{\rho}}\int dx ~\mathcal{H} = A(\rho,r) + \nabla (D(\rho) \nabla \rho - 2\sigma_d(\rho)(\nabla \hat{\rho} + \lambda))
	-  2\sigma_\rho(\rho)\hat{\rho} \\
	\partial_t \hat{\rho} &= - \frac{\delta}{\delta \rho} \int dx~ \mathcal{H} = - \hat{\rho} \frac{\partial A}{\partial \rho} - \hat{r} \frac{\partial B}{\partial \rho} - D(\rho) \nabla^2 \hat{\rho} - \sigma_d'(\rho)(\nabla \hat{\rho} + \lambda)^2 + \sigma_\rho'(\rho)\hat{\rho}^2 \\
	\partial_t r &= \frac{\delta}{\delta \hat{r}} \int dx~ \mathcal{H} = B(\rho,r) - 2\sigma_r(r) \hat{r}\\
	\partial_t \hat{r} &= - \frac{\delta}{\delta r} \int dx~ \mathcal{H} = - \hat{\rho} \frac{\partial A}{\partial r} - \hat{r}\frac{\partial B}{\partial r} + \sigma_r'(r)\hat{r}^2 
\end{align}
From the first and third Hamilton equation for $\rho$ and $r$ respectively, we have the steady states
\begin{align}
	\hat{\rho}_0 &= \frac{A_0}{2\sigma_{\rho,0}} = \frac{ (Z_1 r_0 - \gamma)}{\frac{2}{N}(Z_1r_0 + \gamma)}\label{eq:hatrho}\\
	\hat{r}_0 &= \frac{B_0}{2\sigma_{r,0}} = \frac{1-\frac{r_0}{\kappa} - Z_2 \rho_0}{\frac{2}{\kappa}(1+\frac{r_0}{\kappa} + Z_2 \rho_0)}
\end{align}

Substituting the above relations into the Hamilton-Jacobi equations for $\rho$ and $r$ at steady-state, then $\rho_0$ and $r_0$ are defined by the implicit relations
\begin{align} 
	0 &= - (\hat{\rho}_0)\partial_\rho A_0 - \hat{r}_0 \partial_\rho B_0 - \lambda^2\sigma_{d,0}' + \sigma_{\rho,0}' \hat{\rho}_0^2 \notag \\
        &=  \frac{ (Z_1 r_0 - \gamma)^2}{\frac{4}{N}(Z_1r_0 + \gamma)} +\frac{Z_2r_0\phi(0)(1-\frac{r_0}{\kappa} - Z_2 \rho_0)}{\frac{2}{\kappa}(1+\frac{r_0}{\kappa} + Z_2 \rho_0)} -2\lambda^2 (1- 2\rho_0) \label{eq:rhor_HJ1}  \\
	0 &= - (\hat{\rho}_0 ) \partial_r A_0 - \hat{r}_0 \partial_r B_0 + \sigma'_{r,0} \hat{r}_0^2\notag\\
    &= \frac{ Z_1\rho_0\psi(0)(Z_1 r_0 - \gamma)}{\frac{2}{N}(Z_1r_0 + \gamma)} - \frac{(1-\frac{r_0}{\kappa} - Z_2 \rho_0)(1-\frac{2r_0}{\kappa} - Z_2 \rho_0)}{\frac{2}{\kappa}(1+\frac{r_0}{\kappa} + Z_2 \rho_0)} \notag\\
    &\phantom{==} + \frac{(1-\frac{r_0}{\kappa} - Z_2 \rho_0)^2(1+\frac{2r_0}{\kappa} - Z_2 \rho_0)}{\frac{4}{\kappa}(1+\frac{r_0}{\kappa} + Z_2 \rho_0)^2}\label{eq:rhor_HJ2}
\end{align}
Furthermore, the large deviation function can now be written as 
\begin{align}
    \psi(\lambda) =  \min_{\rho_0,r_0}\left\{ \frac{\rho_0(Z_1r_0 - \gamma)^2}{\frac{4}{N}(Z_1r_0 + \gamma)} + \frac{r_0(1-\frac{r_0}{\kappa}-Z_2\rho_0)^2}{\frac{4}{\kappa}(1-\frac{r_0}{\kappa}-Z_2\rho_0)} + 2\lambda^2 \rho_0(1-\rho_0)\right\}
\end{align}
where we only have to minimize with respect to $\rho_0$ and $r_0$. This is done by solving Eqs. \eqref{eq:rhor_HJ1} and \eqref{eq:rhor_HJ2} for $\rho_0$ and $r_0$ and substituting them into the above equation for $\psi(\lambda)$.

The rate function can be computed by using the Legendre transform
\begin{align}
    I(j) &= \lambda(j) \left[\frac{\partial\psi}{\partial\lambda}\right]_{\lambda=\lambda(j)} - \psi(\lambda(j))\\
        &= \min_{\rho_0,r_0}\left\{- \frac{\rho_0(Z_1r_0 - \gamma)^2}{\frac{4}{N}(Z_1r_0 + \gamma)} - \frac{r_0(1-\frac{r_0}{\kappa}-Z_2\rho_0)^2}{\frac{4}{\kappa}(1-\frac{r_0}{\kappa}-Z_2\rho_0)} + 2\lambda^2 \rho_0(1-\rho_0)\right\}
\end{align}
where $j(\lambda) = \frac{\partial \psi}{\partial \lambda}$.

\section{Expansion About the Mean Densities}
The average densities $\rho_0$ and $r_0$ can be found by solving the remaining Hamilton-Jacobi equations. This represents the homogeneous phase of the system where the ecosystem has a flat density profile across the entire trait space. To investigate the stability of this phase, we need to look at second order variations in the action. 

Let us make an expansion of the action $S$ about the mean densities, i.e. $\rho = \rho_0+ \delta\rho$ and $r = r_0 + \delta r$, up to quadratic order in $\delta \rho$, $\delta r$, $\hat{\rho}$, and $\hat{r}$. Note that we must include $\hat{\rho}$ and $\hat{r}$ because they are near zero around the steady state. Using the expansions of the functions 
\begin{align}
	D(\rho) &= D_0 + D_0'\delta\rho + \frac{D_0''}{2}\delta \rho^2\\
	\sigma_d(\rho) &= \sigma_{d,0} + \sigma'_{d,0} \delta \rho + \frac{\sigma_{d,0}''}{2} \delta\rho^2\\
	\sigma_\rho(\rho) &= \sigma_{\rho,0} \\
	\sigma_r(r) &= \sigma_{r,0} \\
	A(\rho,r) &= \frac{\partial A_0}{\partial \rho}\delta \rho + \frac{\partial A_0}{\partial r} \delta r\\
	B(\rho,r) &= \frac{\partial B_0}{\partial \rho}\delta \rho + \frac{\partial B_0}{\partial r} \delta r
\end{align}
the generating functional can be written as 
\begin{align}
	z(\lambda) &=  \int D[\delta\rho, \hat{\rho},\delta r,\hat{r}]~ \exp\left[-LT\int\int dx dt ~ \hat{\rho}\left(\delta\dot{\rho}- \frac{\partial A_0}{\partial \rho}\delta \rho - \frac{\partial A_0}{\partial r} \delta r \right) \right.\notag\\
	&\phantom{==}  + \hat{r}\left(\delta\dot{r} - \frac{\partial B_0}{\partial \rho}\delta \rho - \frac{\partial B_0}{\partial r} \delta r\right) +  ( \nabla\hat{\rho} + \lambda) \left(D_0 + D_0'\delta\rho + \frac{D_0''}{2}\delta \rho^2\right)\nabla \delta \rho \notag\\
	&\phantom{==} \left. - (\nabla \hat{\rho} + \lambda)^2\left(\sigma_{d,0} + \sigma'_{d,0} \delta \rho + \frac{\sigma_{d,0}''}{2} \delta\rho^2\right) + \sigma_{\rho,0}\hat{\rho}^2 + \sigma_{r,0}\hat{r}^2 \right]\\
	&= \int D[\delta\rho, \hat{\rho},\delta r,\hat{r}]~\exp\left[ -LT\int\int dx dt~ \hat{\rho}\left(\delta\dot{\rho} - \frac{\partial A_0}{\partial \rho} \delta \rho - \frac{\partial A_0}{\partial r}\delta r\right) \right.\notag\\
	&\phantom{==} +  \hat{r} \left( \delta \dot{r}- \frac{\partial B_0}{\partial \rho}\delta \rho - \frac{\partial B_0}{\partial r} \delta r\right)+ D_0\nabla \hat{\rho} \nabla\delta \rho + \lambda D_0 \nabla \delta \rho + \lambda D_0' \delta \rho \nabla\delta \rho\notag\\
	&\phantom{==} - \sigma_{d,0}(\nabla \hat{\rho} )^2 - 2\lambda  \nabla \hat{\rho} (\sigma_{d,0} + \sigma_{d,0}'\delta \rho) - \lambda^2 \left(\sigma_{d,0} + \sigma_{d,0}'\delta\rho + \frac{\sigma_{d,0}''}{2} \delta \rho^2 \right) \notag\\
	&\phantom{==} ++ \sigma_{\rho,0}\hat{\rho}^2 + \sigma_{r,0} \hat{r}^2 \bigg]
\end{align}
We can use integration by parts to turn terms like $D_0\nabla \hat{\rho} \nabla \delta \rho$ into $-D_0\hat{\rho} \nabla^2 \delta \rho $ and $2\lambda \sigma_{d,0}' \nabla \hat{\rho} \delta \rho$ into $2\lambda \sigma_{d,0}' \hat{\rho} \nabla \delta \rho$. In addition, if we assume periodic boundary conditions, then we can eliminate terms that are purely gradients of functions of $\rho, \hat{\rho}, r,$ and $\hat{r}$ since integrating them with respect to space yields zero. This allows us to remove terms like $2\lambda \sigma_{d,0} \nabla \hat{\rho}$ and $\lambda D_0' \delta \rho \nabla \delta \rho$, the latter of which can be written as $\frac{1}{2}\lambda D_0' \nabla (\delta\rho^2)$. 

As a result, we are left with 
\begin{align}
	z(\lambda) &= \int  D[\delta\rho, \hat{\rho},\delta r,\hat{r}] ~\exp\left[ -LT\int\int dx dt~ \hat{\rho}\left(\partial_t - D_0 \nabla^2 -2\lambda \sigma_{d,0}' \nabla - \frac{\partial A_0}{\partial \rho}\right)\delta \rho\right.\notag\\
	&\phantom{=} -  \hat{\rho} \frac{\partial A_0}{\partial r}\delta r  + \hat{r} \left( \delta \dot{r}- \frac{\partial B_0}{\partial \rho}\delta \rho - \frac{\partial B_0}{\partial r} \delta r\right) - \sigma_{d,0}(\nabla \hat{\rho} )^2 \notag\\
	&\phantom{=} - \lambda^2 \left(\sigma_{d,0} + \sigma_{d,0}'\delta\rho + \frac{\sigma_{d,0}''}{2} \delta \rho^2 \right) + \sigma_{\rho,0}\hat{\rho}^2 + \sigma_{r,0} \hat{r}^2 \bigg]
\end{align}

\section{Second Variation}
\label{secondV2}
Starting from the general form of the generating function, we substitute the explicit expressions for the functions $A(\rho,r)$, $B(\rho,r)$, the demographic noise terms $\sigma_\rho$ and $\sigma_r$, and the diffusion noise term $\sigma_d$. In addition, we eliminate the momentum variables $\hat{\rho}_0$ and $\hat{r}_0$ using the solutions from the Hamilton-Jacobi equations. The following is a summary of these expressions:
\begin{align}
	A_0 &= \rho_0 (Z_1 r_0 - \gamma)\\
	B_0 &= r_0 \left(1- \frac{r_0}{K} - Z_2 \rho_0\right) \\
	\sigma_{\rho} &= \frac{1}{N_c}\rho(Z_1\psi\star r  + \gamma) \\
	\sigma_{r} &= \frac{1}{\kappa} r\left(1 + \frac{r}{K} + Z_2 \phi\star \rho\right)\\
	\sigma_{d} &= 2\rho(1-\rho)\\
	\hat{\rho}_0 &= \frac{A_0}{2\sigma_{\rho,0}} =  \frac{\rho_0 (Z_1 r_0 - \gamma) }{\frac{2}{N_c}\rho_0(Z_1 r_0 + \gamma)}\\
\hat{r}_0 &= \frac{B_0}{2\sigma_{r,0}} = \frac{r_0\left(1-\frac{r_0}{K} - Z_2\rho_0\right)}{\frac{2}{\kappa}r_0 \left(1 + \frac{r_0}{K} + Z_2 \rho_0\right)}
\end{align} 
From this, we can evaluate each term in the second order variation expansion of the action in terms of the ecosystem parameters. Because of the Wick rotation for the momenta variables, we must perturb them in the imaginary direction, i.e. $\hat{\rho} \approx \hat{\rho}_0 + i \delta \hat{\rho}$ and $ \hat{r} \approx \hat{r}_0 + i \delta \hat{r}$

\underline{Term 1:}
\begin{align}
\hat{\rho} ( \dot{\rho} - A(\rho,r)) &\rightarrow 	(\hat{\rho}_0 + i \delta \hat{\rho}) \left((\dot{\rho}_0 + \dot{\delta\rho} ) -A(\rho_0 + \delta \rho ,r_0 + \delta r)\right)\notag\\
 	&\rightarrow i\delta \hat{\rho} \left(\dot{\delta \rho} - (Z_1 r_0 - \gamma) \delta \rho - \rho_0 (Z_1 \psi\star \delta r) \right) + \hat{\rho}_0 (-Z_1 \delta \rho \psi \star \delta r)\notag\\
 	&\rightarrow i \delta \hat{\rho}_{-n,-m} \left( i\omega_m  \delta\rho_{n,m} - (Z_1r_0 - \gamma)\delta\rho_{n,m} - \rho_0 Z_1 \psi(q_n) \delta r_{n,m} )  \right) \notag\\
  &\phantom{==} - Z_1 \hat{\rho_0} \psi(q_n) \delta \rho_{-n,-m} \delta r_{n,m}
\end{align}
\underline{Term 2:}
\begin{align}
	\hat{r} (\dot{r}-B(\rho,r)) &\rightarrow (\hat{r}_0 + i\delta \hat{r}) \left((\dot{r}_0 + \dot{\delta r} ) - B(\rho_0 + \delta \rho , r_0 + \delta r)\right)\notag\\
	&\rightarrow i\delta \hat{r} \left(\dot{\delta r} - (-Z_2r_0 \phi\star \delta \rho) - \left(1- 2\frac{r_0}{K} - Z_2\rho_0 \right)\delta r \right) + \hat{r}_0 \left(Z_2 \delta r \phi\star \delta \rho + \frac{\delta r^2}{K}\right)\notag\\
	&\rightarrow i \delta \hat{r}_{-n,-m} \left( i\omega_m  \delta r_{n,m}  + Z_2r_0 \phi(q_n)\rho_{n,m} - \left(1- 2\frac{r_0}{K} - Z_2\rho_0 \right)\delta r_{n,m} \right) \notag\\
	&\phantom{\rightarrow} + \left( Z_2 \hat{r}_0 \phi(q_n) \delta r_{-n,-m} \delta \rho_{n,m} + \frac{\hat{r_0}}{K}|\delta r_{n,m}|^2\right)\notag\\
\end{align}
\underline{Term 3:}
\begin{align}
	(\nabla \hat{\rho} + \lambda) (D(\rho) \nabla \rho) ) &\rightarrow \left(i\nabla \delta \hat{\rho} + \lambda \right)\left(D_0 + D'_0 \delta \rho + \frac{1}{2} D''_0 \delta \rho^2\right)\left(\nabla \delta \rho\right)\\
	&\rightarrow   iD_0 (\nabla \delta \hat{\rho})( \nabla \delta \rho) + \lambda D_0' \delta \rho \nabla \delta \rho \\
	&\rightarrow - i D_0 \delta \hat{\rho} \nabla^2\delta \rho + \frac{1}{2}\lambda D_0' \nabla \delta \rho^2\\
	&\rightarrow iD_0 q_n^2\delta \hat{\rho}_{-n,-m}\delta \rho_{n,m}
\end{align}
\underline{Term 4:}
\begin{align}
	- (\nabla \hat{\rho} + \lambda)^2 \sigma_d(\rho) &\rightarrow -(i\nabla \delta\hat{\rho} + \lambda)^2 \left(\sigma_{d,0} + \sigma_{d,0}' \delta\rho + \frac{1}{2} \sigma_{d,0}'' \delta \rho^2\right)\notag\\
	&\rightarrow \sigma_{d,0} (\nabla \delta \hat{\rho})^2 - 2i\lambda \sigma'_{d,0} (\nabla \delta\hat{\rho}) \delta \rho + \frac{\lambda^2}{2} \sigma_{d,0}'' \delta \rho^2 \notag\\
	&\rightarrow  \sigma_{d,0} (\nabla \delta \hat{\rho})^2 + 2i\lambda \sigma'_{d,0}  \delta\hat{\rho}(\nabla\delta \rho) + \frac{\lambda^2}{2} \sigma_{d,0}'' \delta \rho^2 \notag\\
	&\rightarrow  \sigma_{d,0} q^2_n |\delta \hat{\rho}_{n,m}|^2 -2i\lambda \sigma_{d,0}'q_n \delta \hat{\rho}_{-n,-m} \delta \rho_{n,m} + \frac{\lambda^2}{2} \sigma_{d,0}'' |\delta\rho_{n,m}|^2 
\end{align}
\underline{Term 5:}
\begin{align}
	\sigma_{\rho}\hat{\rho}^2 &\rightarrow \left( \sigma_{\rho,0} + \frac{1}{N_c} \delta\rho \left( Z_1 r_0 + \gamma\right) + \frac{1}{N_c} Z_1\rho_0(\psi \star \delta r)  + \frac{1}{N_c}Z_1 \delta \rho \psi \star \delta r \right)\notag\\
 &\phantom{==} \times\left(\hat{\rho}_0^2 + 2i\hat{\rho}_0\delta \hat{\rho} - \delta \hat{\rho}_0^2  \right)\notag\\
	&\rightarrow -\sigma_{\rho,0} \delta \hat{\rho}^2 +i\frac{2\hat{\rho}_0}{N_c} \delta \hat{\rho} \left[(Z_1r_0 + \gamma) \delta \rho + Z_1 \rho_0 \psi \star \delta r\right] + \frac{\hat{\rho}_0^2}{N_c}Z_1 \delta \rho \psi\star \delta r \notag\\
	&\rightarrow -\sigma_{\rho,0} |\delta \hat{\rho}_{n,m}|^2 + i\frac{2\hat{\rho}_0}{N_c} \delta \hat{\rho}_{-n,-m} \left[ (Z_1 r_0 + \gamma) \delta \rho_{n,m} + Z_1 \rho_0 \psi(q_n) \delta r_{n,m}\right] \notag\\
 &\phantom{==}+ \frac{\hat{\rho}_0^2}{N_c}Z_1 \psi(q_n) \delta \rho_{-n,-m} \delta r_{n,m}
\end{align}
\underline{Term 6:}
\begin{align}
	\sigma_{r}\hat{r}^2 &\rightarrow \left( \sigma_{r,0} + \frac{1}{\kappa} \delta r \left( 1 + \frac{2r_0}{K}+ Z_2 \rho_0\right)  + \frac{1}{\kappa} Z_2 r_0(\phi \star \delta \rho)  + \frac{1}{\kappa}Z_2 \delta r \phi \star \delta \rho \right)\notag\\
 &\phantom{==}\times\left(\hat{r}_0 + 2i\hat{r}_0\delta \hat{r} - \delta \hat{r}_0^2 \right)\notag\\
	&\rightarrow -\sigma_{r,0} \delta \hat{r}^2 +i\frac{2\hat{r}_0}{\kappa} \delta \hat{r} \left[\left(1 + \frac{2r_0}{K} + Z_2 \rho_0\right)\delta r + Z_2 r_0 \phi \star \delta \rho \right] + \frac{\hat{r}_0^2}{\kappa} Z_2 \delta r \phi \star \delta \rho \notag\\
	&\rightarrow -\sigma_{r,0} |\delta \hat{r}_{n,m}|^2 +i\frac{2\hat{r}_0}{\kappa} \delta \hat{r}_{-n,-m} \left[\left(1 + \frac{2r_0}{K} + Z_2 \rho_0\right)\delta r_{n,m} + Z_2 r_0 \phi(q_n) \delta \rho_{n,m} \right] \notag\\
	&\phantom{\rightarrow} + \frac{\hat{r}_0^2}{\kappa} Z_2 \phi(q_n)\delta r_{-n,-m}  \delta \rho_{n,m}
\end{align}

Collecting similar terms together, we can write the quadratic variation in the action as
\[
	\delta S = L T \sum_{n,m} V^T_{-n,-m}C_{n,m} V_{n,m}
\]
where 
\begin{align}
	V_{n,m} = \begin{pmatrix}
			\delta \rho^{n,m}\\
			\delta p_{\rho}^{n,m}\\
			\delta r^{n,m}\\
			\delta p_{r}^{n,m}
		\end{pmatrix} \qquad, \qquad q_n = \frac{2\pi n}{L} \qquad, \qquad \omega_m = \frac{2\pi m}{T}
\end{align}

\begin{align}
C^{n,m} &= \begin{pmatrix}
	C_{\rho,\rho} & C_{\rho,\hat{\rho}} & C_{\rho,r} & C_{\rho,\hat{r}}\\
	C_{\hat{\rho},\rho} & C_{\hat{\rho},\hat{\rho}} & C_{\hat{\rho},r} & C_{\hat{\rho},\hat{r}}\\
	C_{r,\rho} & C_{r,\hat{\rho}} & C_{r,r} & C_{r,\hat{r}}\\
	C_{\hat{r},\rho} & C_{\hat{r},\hat{\rho}} & C_{\hat{r},r} & C_{\hat{r},\hat{r}}
\end{pmatrix}
\end{align}
\begin{align}
	C_{\rho,\rho} &= -\frac{\lambda^2\sigma_{d,0}'' }{2}\\
	C_{\rho,\hat{\rho}} &= \frac{1}{2}\left(\omega_m + 2\lambda\sigma_{d,0}'q_n + i\left(D_0 q_n^2 - (Z_1r_0 - \gamma) + \frac{2\hat{\rho}}{N_c}(Z_1r_0 + \gamma)\right)\right)\\
	C_{\rho,r} &= \frac{1}{2} \left(Z_2\hat{r}_0 \phi(-q_n)- Z_1 \hat{\rho_0} \psi(q_n) + \frac{\hat{\rho}_0^2}{N_c}Z_1 \psi(q_n) + \frac{\hat{r}_0^2}{\kappa} Z_2 \phi(-q_n) \right)\\
	C_{\rho,\hat{r}} &= \frac{i}{2}\left( Z_2r_0 \phi(-q_n) + \frac{2\hat{r}_0}{\kappa} Z_2 r_0 \phi(-q_n)\right) \\
	C_{\hat{\rho},\rho} &=  \frac{1}{2}\left(-\omega_m - 2\lambda\sigma_{d,0}'q_n + i\left(D_0 q_n^2 - (Z_1r_0 - \gamma) + \frac{2\hat{\rho}}{N_c}(Z_1r_0 + \gamma)\right)\right)\\
	C_{\hat{\rho},\hat{\rho}} &= \sigma_{d,0}q_n^2 - \sigma_{\rho,0}\\
	C_{\hat{\rho},r} &= \frac{i}{2}\left(- Z_1\rho_0 \psi(q_n) +  \frac{2\hat{\rho}_0}{N_c}Z_1 \rho_0 \psi(q_n) \right)\\
	C_{\hat{\rho},\hat{r}} &= 0\\
	C_{r,\rho} &=  \frac{1}{2} \left(Z_2\hat{r}_0 \phi(q_n)- Z_1 \hat{\rho_0} \psi(-q_n) + \frac{\hat{\rho}_0^2}{N_c}Z_1 \psi(-q_n) + \frac{\hat{r}_0^2}{\kappa} Z_2 \phi(q_n)\right)  \\
	C_{r,\hat{\rho}} &=  \frac{i}{2}\left(- Z_1\rho_0 \psi(-q_n) +  \frac{2\hat{\rho}_0}{N_c}Z_1 \rho_0 \psi(-q_n) \right)\\
	C_{r,r} &= \frac{\hat{r}_0}{K}\\
	C_{r,\hat{r}}&= \frac{1}{2}\left(\omega_m -  i\left(1 - 2\frac{r_0}{K} - Z_2\rho_0\right)+ i \frac{2\hat{r}_0}{\kappa}\left(1 + \frac{2r_0}{K} + Z_2 \rho_0\right)\right) \\
	C_{\hat{r},\rho} &=  \frac{i}{2}\left( Z_2r_0 \phi(q_n) + \frac{2\hat{r}_0}{\kappa} Z_2 r_0 \phi(q_n)\right) \\
	C_{\hat{r},\hat{\rho}} &= 0\\
	C_{\hat{r},r} &= \frac{1}{2}\left(-\omega_m -  i\left(1 - 2\frac{r_0}{K} - Z_2\rho_0\right) + i \frac{2\hat{r}_0}{\kappa}\left(1 + \frac{2r_0}{K} + Z_2 \rho_0\right)\right)   \\
	C_{\hat{r},\hat{r}} &= -\sigma_{r,0}  
\end{align}

The homogeneous solution is stable only when the matrix $C_{n,m}$ is positive definite. However, if one of the eigenvalues becomes negative for some mode $(n,m)$, then the homogeneous solution is no longer stable and the profiles $\rho(x,t)$ and $r(x,t)$ become spatially modulated and/or time-dependent. To find the parameters that trigger this condition, we first note that higher values of $n$ incur a larger cost in the action due to more spatial modulations. This indicates that the most unstable spatial mode is the $n=1$ mode, or $q^* = 2\pi/\ell$. So, to find the locally unstable region, we just have to find the temporal mode $\omega_m$ such that makes the matrix no longer positive definite. 

For the homogeneous state to be stable, the $C_{n,m}$ matrix must be positive-definite. This implies that the determinant of $C_{n,m}$ is positive for all values of $n,m$. If one of the eigenvalues becomes negative for any interval of spatial frequencies $k_n$ and/or temporal frequencies $\omega_m$, then homogeneous state becomes unstable and the system exhibits traveling waves. The boundary between the homogeneous state and the traveling wave state is set by the conditions
\begin{align}
	\det(C_{n,m}(\omega,q=q_*)) &= 0 \\
 \partial_\omega\det(C_{n,m}(\omega,q=q_*)) &= 0
\end{align}
The first equation sets the determinant of $C_{n,m}$ equal to zero, which is used to determine the mode $\omega_m$ that makes the homogeneous solution unstable. The second equation ensures that this $\omega_m$ corresponds to a minimum of the determinant, making it the most unstable mode that sets the temporal scale of the consumer and resource dynamics.

The determinant of this matrix is a fourth order polynomial in $\omega$ with coefficients that are complicated functions of the model parameters and the wavenumber $q$. In the large $L$ limit, the determinant reduces to a quartic of the form $\frac{1}{16}\omega^4 + \frac{b}{8} \omega^2 + c$.  Taking the derivative with respect to $\omega$ yields $\frac{1}{4}\omega^3 + \frac{b}{4}\omega$, which has roots at $\omega = 0$ and $\pm \sqrt{b}$. These are the extrema of the quartic. If $b>0$, then the minimum is at $\omega = 0$ and the homogeneous solution is unstable if 
\begin{align}
    c < 0
\end{align}
On the other hand, if $b<0$, then the minima of the quartic polynomial are $\pm \sqrt{-b}$. Substituting this back into the quartic polynomial, we find that the determinant is negative when
\begin{align}
    \frac{3}{16}b^2 + c < 0
\end{align}


\begin{thebibliography}{85}%
\makeatletter
\providecommand \@ifxundefined [1]{%
 \@ifx{#1\undefined}
}%
\providecommand \@ifnum [1]{%
 \ifnum #1\expandafter \@firstoftwo
 \else \expandafter \@secondoftwo
 \fi
}%
\providecommand \@ifx [1]{%
 \ifx #1\expandafter \@firstoftwo
 \else \expandafter \@secondoftwo
 \fi
}%
\providecommand \natexlab [1]{#1}%
\providecommand \enquote  [1]{``#1''}%
\providecommand \bibnamefont  [1]{#1}%
\providecommand \bibfnamefont [1]{#1}%
\providecommand \citenamefont [1]{#1}%
\providecommand \href@noop [0]{\@secondoftwo}%
\providecommand \href [0]{\begingroup \@sanitize@url \@href}%
\providecommand \@href[1]{\@@startlink{#1}\@@href}%
\providecommand \@@href[1]{\endgroup#1\@@endlink}%
\providecommand \@sanitize@url [0]{\catcode `\\12\catcode `\$12\catcode
  `\&12\catcode `\#12\catcode `\^12\catcode `\_12\catcode `\%12\relax}%
\providecommand \@@startlink[1]{}%
\providecommand \@@endlink[0]{}%
\providecommand \url  [0]{\begingroup\@sanitize@url \@url }%
\providecommand \@url [1]{\endgroup\@href {#1}{\urlprefix }}%
\providecommand \urlprefix  [0]{URL }%
\providecommand \Eprint [0]{\href }%
\providecommand \doibase [0]{http://dx.doi.org/}%
\providecommand \selectlanguage [0]{\@gobble}%
\providecommand \bibinfo  [0]{\@secondoftwo}%
\providecommand \bibfield  [0]{\@secondoftwo}%
\providecommand \translation [1]{[#1]}%
\providecommand \BibitemOpen [0]{}%
\providecommand \bibitemStop [0]{}%
\providecommand \bibitemNoStop [0]{.\EOS\space}%
\providecommand \EOS [0]{\spacefactor3000\relax}%
\providecommand \BibitemShut  [1]{\csname bibitem#1\endcsname}%
\let\auto@bib@innerbib\@empty
\bibitem [{\citenamefont {Lotka}(1920)}]{lotkaAnalyticalNoteCertain1920}%
  \BibitemOpen
  \bibfield  {author} {\bibinfo {author} {\bibfnamefont {A.~J.}\ \bibnamefont
  {Lotka}},\ }\href {\doibase 10.1073/pnas.6.7.410} {\bibfield  {journal}
  {\bibinfo  {journal} {Proceedings of the National Academy of Sciences}\
  }\textbf {\bibinfo {volume} {6}},\ \bibinfo {pages} {410} (\bibinfo {year}
  {1920})}\BibitemShut {NoStop}%
\bibitem [{\citenamefont {Lotka}(1970)}]{lotkaElementsMathematicalBiology1970}%
  \BibitemOpen
  \bibfield  {author} {\bibinfo {author} {\bibfnamefont {A.~J.}\ \bibnamefont
  {Lotka}},\ }\href@noop {} {{\selectlanguage {english}\emph {\bibinfo {title}
  {Elements of mathematical biology}}}},\ \bibinfo {edition} {unabrigded
  republ}\ ed.\ (\bibinfo  {publisher} {Dover},\ \bibinfo {address} {New
  York},\ \bibinfo {year} {1970})\BibitemShut {NoStop}%
\bibitem [{\citenamefont
  {Volterra}(1926{\natexlab{a}})}]{volterraVariazioniFluttuazioniNumero1926}%
  \BibitemOpen
  \bibfield  {author} {\bibinfo {author} {\bibfnamefont {V.}~\bibnamefont
  {Volterra}},\ }\href@noop {} {\bibfield  {journal} {\bibinfo  {journal}
  {Memoire della R. Accademia Nazionale dei Lincei}\ ,\ \bibinfo {pages} {31}}
  (\bibinfo {year} {1926}{\natexlab{a}})}\BibitemShut {NoStop}%
\bibitem [{\citenamefont
  {Volterra}(1926{\natexlab{b}})}]{volterraFluctuationsAbundanceSpecies1926}%
  \BibitemOpen
  \bibfield  {author} {\bibinfo {author} {\bibfnamefont {V.}~\bibnamefont
  {Volterra}},\ }\href {\doibase 10.1038/118558a0} {\bibfield  {journal}
  {\bibinfo  {journal} {Nature}\ }\textbf {\bibinfo {volume} {118}},\ \bibinfo
  {pages} {558} (\bibinfo {year} {1926}{\natexlab{b}})}\BibitemShut {NoStop}%
\bibitem [{\citenamefont {Macarthur}\ and\ \citenamefont
  {Levins}(1967)}]{macarthurLimitingSimilarityConvergence1967}%
  \BibitemOpen
  \bibfield  {author} {\bibinfo {author} {\bibfnamefont {R.}~\bibnamefont
  {Macarthur}}\ and\ \bibinfo {author} {\bibfnamefont {R.}~\bibnamefont
  {Levins}},\ }\href {\doibase 10.1086/282505} {\bibfield  {journal} {\bibinfo
  {journal} {The American Naturalist}\ }\textbf {\bibinfo {volume} {101}},\
  \bibinfo {pages} {377} (\bibinfo {year} {1967})}\BibitemShut {NoStop}%
\bibitem [{\citenamefont
  {MacArthur}(1970{\natexlab{a}})}]{macarthurSpeciesPackingCompetitive1970}%
  \BibitemOpen
  \bibfield  {author} {\bibinfo {author} {\bibfnamefont {R.}~\bibnamefont
  {MacArthur}},\ }\href {\doibase 10.1016/0040-5809(70)90039-0} {\bibfield
  {journal} {\bibinfo  {journal} {Theoretical Population Biology}\ }\textbf
  {\bibinfo {volume} {1}},\ \bibinfo {pages} {1} (\bibinfo {year}
  {1970}{\natexlab{a}})}\BibitemShut {NoStop}%
\bibitem [{\citenamefont
  {Chesson}(1990)}]{chessonMacArthurConsumerresourceModel1990}%
  \BibitemOpen
  \bibfield  {author} {\bibinfo {author} {\bibfnamefont {P.}~\bibnamefont
  {Chesson}},\ }\href {\doibase 10.1016/0040-5809(90)90025-Q} {\bibfield
  {journal} {\bibinfo  {journal} {Theoretical Population Biology}\ }\textbf
  {\bibinfo {volume} {37}},\ \bibinfo {pages} {26} (\bibinfo {year}
  {1990})}\BibitemShut {NoStop}%
\bibitem [{\citenamefont
  {Kimura}(1955)}]{kimuraStochasticProcessesDistribution1955}%
  \BibitemOpen
  \bibfield  {author} {\bibinfo {author} {\bibfnamefont {M.}~\bibnamefont
  {Kimura}},\ }\href {\doibase 10.1101/SQB.1955.020.01.006} {\bibfield
  {journal} {\bibinfo  {journal} {Cold Spring Harbor Symposia on Quantitative
  Biology}\ }\textbf {\bibinfo {volume} {20}},\ \bibinfo {pages} {33} (\bibinfo
  {year} {1955})}\BibitemShut {NoStop}%
\bibitem [{\citenamefont
  {Gillespie}(2000)}]{gillespieGeneticDriftInfinite2000}%
  \BibitemOpen
  \bibfield  {author} {\bibinfo {author} {\bibfnamefont {J.~H.}\ \bibnamefont
  {Gillespie}},\ }\href {\doibase 10.1093/genetics/155.2.909} {\bibfield
  {journal} {\bibinfo  {journal} {Genetics}\ }\textbf {\bibinfo {volume}
  {155}},\ \bibinfo {pages} {909} (\bibinfo {year} {2000})}\BibitemShut
  {NoStop}%
\bibitem [{\citenamefont {Neher}\ and\ \citenamefont
  {Walczak}(2018)}]{neherProgressOpenProblems2018}%
  \BibitemOpen
  \bibfield  {author} {\bibinfo {author} {\bibfnamefont {R.~A.}\ \bibnamefont
  {Neher}}\ and\ \bibinfo {author} {\bibfnamefont {A.~M.}\ \bibnamefont
  {Walczak}},\ }\href {http://arxiv.org/abs/1804.07720} {\bibfield  {journal}
  {\bibinfo  {journal} {arXiv:1804.07720 [q-bio]}\ } (\bibinfo {year}
  {2018})},\ \bibinfo {note} {arXiv: 1804.07720}\BibitemShut {NoStop}%
\bibitem [{\citenamefont {Doebeli}(2011)}]{doebeliAdaptiveDiversification2011}%
  \BibitemOpen
  \bibfield  {author} {\bibinfo {author} {\bibfnamefont {M.}~\bibnamefont
  {Doebeli}},\ }\href@noop {} {\emph {\bibinfo {title} {Adaptive
  diversification}}},\ \bibinfo {series} {Monographs in population biology}\
  No.~\bibinfo {number} {48}\ (\bibinfo  {publisher} {Princeton University
  Press},\ \bibinfo {address} {Princeton, N.J},\ \bibinfo {year} {2011})\
  \bibinfo {note} {oCLC: ocn687685579}\BibitemShut {NoStop}%
\bibitem [{\citenamefont {Biller}\ \emph {et~al.}(2015)\citenamefont {Biller},
  \citenamefont {Berube}, \citenamefont {Lindell},\ and\ \citenamefont
  {Chisholm}}]{billerProchlorococcusStructureFunction2015}%
  \BibitemOpen
  \bibfield  {author} {\bibinfo {author} {\bibfnamefont {S.~J.}\ \bibnamefont
  {Biller}}, \bibinfo {author} {\bibfnamefont {P.~M.}\ \bibnamefont {Berube}},
  \bibinfo {author} {\bibfnamefont {D.}~\bibnamefont {Lindell}}, \ and\
  \bibinfo {author} {\bibfnamefont {S.~W.}\ \bibnamefont {Chisholm}},\ }\href
  {\doibase 10.1038/nrmicro3378} {\bibfield  {journal} {\bibinfo  {journal}
  {Nature Reviews Microbiology}\ }\textbf {\bibinfo {volume} {13}},\ \bibinfo
  {pages} {13} (\bibinfo {year} {2015})}\BibitemShut {NoStop}%
\bibitem [{\citenamefont {Braakman}\ \emph {et~al.}(2017)\citenamefont
  {Braakman}, \citenamefont {Follows},\ and\ \citenamefont
  {Chisholm}}]{braakmanMetabolicEvolutionSelforganization2017}%
  \BibitemOpen
  \bibfield  {author} {\bibinfo {author} {\bibfnamefont {R.}~\bibnamefont
  {Braakman}}, \bibinfo {author} {\bibfnamefont {M.~J.}\ \bibnamefont
  {Follows}}, \ and\ \bibinfo {author} {\bibfnamefont {S.~W.}\ \bibnamefont
  {Chisholm}},\ }\href {\doibase 10.1073/pnas.1619573114} {\bibfield  {journal}
  {\bibinfo  {journal} {Proceedings of the National Academy of Sciences}\
  }\textbf {\bibinfo {volume} {114}} (\bibinfo {year} {2017}),\
  10.1073/pnas.1619573114}\BibitemShut {NoStop}%
\bibitem [{\citenamefont {Rosen}\ \emph {et~al.}(2015)\citenamefont {Rosen},
  \citenamefont {Davison}, \citenamefont {Bhaya},\ and\ \citenamefont
  {Fisher}}]{rosenFinescaleDiversityExtensive2015}%
  \BibitemOpen
  \bibfield  {author} {\bibinfo {author} {\bibfnamefont {M.~J.}\ \bibnamefont
  {Rosen}}, \bibinfo {author} {\bibfnamefont {M.}~\bibnamefont {Davison}},
  \bibinfo {author} {\bibfnamefont {D.}~\bibnamefont {Bhaya}}, \ and\ \bibinfo
  {author} {\bibfnamefont {D.~S.}\ \bibnamefont {Fisher}},\ }\href {\doibase
  10.1126/science.aaa4456} {\bibfield  {journal} {\bibinfo  {journal}
  {Science}\ }\textbf {\bibinfo {volume} {348}},\ \bibinfo {pages} {1019}
  (\bibinfo {year} {2015})}\BibitemShut {NoStop}%
\bibitem [{\citenamefont {Hanski}\ and\ \citenamefont
  {Saccheri}(2006)}]{hanskiMolecularLevelVariationAffects2006}%
  \BibitemOpen
  \bibfield  {author} {\bibinfo {author} {\bibfnamefont {I.}~\bibnamefont
  {Hanski}}\ and\ \bibinfo {author} {\bibfnamefont {I.}~\bibnamefont
  {Saccheri}},\ }\href {\doibase 10.1371/journal.pbio.0040129} {\bibfield
  {journal} {\bibinfo  {journal} {PLoS Biology}\ }\textbf {\bibinfo {volume}
  {4}},\ \bibinfo {pages} {e129} (\bibinfo {year} {2006})}\BibitemShut
  {NoStop}%
\bibitem [{\citenamefont {Zheng}\ \emph {et~al.}(2009)\citenamefont {Zheng},
  \citenamefont {Ovaskainen},\ and\ \citenamefont
  {Hanski}}]{zhengModellingSingleNucleotide2009}%
  \BibitemOpen
  \bibfield  {author} {\bibinfo {author} {\bibfnamefont {C.}~\bibnamefont
  {Zheng}}, \bibinfo {author} {\bibfnamefont {O.}~\bibnamefont {Ovaskainen}}, \
  and\ \bibinfo {author} {\bibfnamefont {I.}~\bibnamefont {Hanski}},\ }\href
  {\doibase 10.1098/rstb.2009.0005} {\bibfield  {journal} {\bibinfo  {journal}
  {Philosophical Transactions of the Royal Society B: Biological Sciences}\
  }\textbf {\bibinfo {volume} {364}},\ \bibinfo {pages} {1519} (\bibinfo {year}
  {2009})}\BibitemShut {NoStop}%
\bibitem [{\citenamefont {Palkovacs}\ and\ \citenamefont
  {Post}(2008)}]{palkovacsEcoevolutionaryInteractionsPredators2008}%
  \BibitemOpen
  \bibfield  {author} {\bibinfo {author} {\bibfnamefont {E.~P.}\ \bibnamefont
  {Palkovacs}}\ and\ \bibinfo {author} {\bibfnamefont {D.~M.}\ \bibnamefont
  {Post}},\ }\href@noop {} {\bibfield  {journal} {\bibinfo  {journal}
  {Evolutionary Ecology Research}\ }\textbf {\bibinfo {volume} {10}},\ \bibinfo
  {pages} {699} (\bibinfo {year} {2008})}\BibitemShut {NoStop}%
\bibitem [{\citenamefont {Palkovacs}\ and\ \citenamefont
  {Post}(2009)}]{palkovacsExperimentalEvidenceThat2009}%
  \BibitemOpen
  \bibfield  {author} {\bibinfo {author} {\bibfnamefont {E.~P.}\ \bibnamefont
  {Palkovacs}}\ and\ \bibinfo {author} {\bibfnamefont {D.~M.}\ \bibnamefont
  {Post}},\ }\href {\doibase 10.1890/08-1673.1} {\bibfield  {journal} {\bibinfo
   {journal} {Ecology}\ }\textbf {\bibinfo {volume} {90}},\ \bibinfo {pages}
  {300} (\bibinfo {year} {2009})}\BibitemShut {NoStop}%
\bibitem [{\citenamefont {Post}\ \emph {et~al.}(2008)\citenamefont {Post},
  \citenamefont {Palkovacs}, \citenamefont {Schielke},\ and\ \citenamefont
  {Dodson}}]{postIntraspecificVariationPredator2008}%
  \BibitemOpen
  \bibfield  {author} {\bibinfo {author} {\bibfnamefont {D.~M.}\ \bibnamefont
  {Post}}, \bibinfo {author} {\bibfnamefont {E.~P.}\ \bibnamefont {Palkovacs}},
  \bibinfo {author} {\bibfnamefont {E.~G.}\ \bibnamefont {Schielke}}, \ and\
  \bibinfo {author} {\bibfnamefont {S.~I.}\ \bibnamefont {Dodson}},\ }\href
  {\doibase 10.1890/07-1216.1} {\bibfield  {journal} {\bibinfo  {journal}
  {Ecology}\ }\textbf {\bibinfo {volume} {89}},\ \bibinfo {pages} {2019}
  (\bibinfo {year} {2008})}\BibitemShut {NoStop}%
\bibitem [{\citenamefont {Schweitzer}\ \emph
  {et~al.}(2008{\natexlab{a}})\citenamefont {Schweitzer}, \citenamefont
  {Madritch}, \citenamefont {Bailey}, \citenamefont {LeRoy}, \citenamefont
  {Fischer}, \citenamefont {Rehill}, \citenamefont {Lindroth}, \citenamefont
  {Hagerman}, \citenamefont {Wooley}, \citenamefont {Hart},\ and\ \citenamefont
  {Whitham}}]{schweitzerGenesEcosystemsGenetic2008}%
  \BibitemOpen
  \bibfield  {author} {\bibinfo {author} {\bibfnamefont {J.~A.}\ \bibnamefont
  {Schweitzer}}, \bibinfo {author} {\bibfnamefont {M.~D.}\ \bibnamefont
  {Madritch}}, \bibinfo {author} {\bibfnamefont {J.~K.}\ \bibnamefont
  {Bailey}}, \bibinfo {author} {\bibfnamefont {C.~J.}\ \bibnamefont {LeRoy}},
  \bibinfo {author} {\bibfnamefont {D.~G.}\ \bibnamefont {Fischer}}, \bibinfo
  {author} {\bibfnamefont {B.~J.}\ \bibnamefont {Rehill}}, \bibinfo {author}
  {\bibfnamefont {R.~L.}\ \bibnamefont {Lindroth}}, \bibinfo {author}
  {\bibfnamefont {A.~E.}\ \bibnamefont {Hagerman}}, \bibinfo {author}
  {\bibfnamefont {S.~C.}\ \bibnamefont {Wooley}}, \bibinfo {author}
  {\bibfnamefont {S.~C.}\ \bibnamefont {Hart}}, \ and\ \bibinfo {author}
  {\bibfnamefont {T.~G.}\ \bibnamefont {Whitham}},\ }\href {\doibase
  10.1007/s10021-008-9173-9} {\bibfield  {journal} {\bibinfo  {journal}
  {Ecosystems}\ }\textbf {\bibinfo {volume} {11}},\ \bibinfo {pages} {1005}
  (\bibinfo {year} {2008}{\natexlab{a}})}\BibitemShut {NoStop}%
\bibitem [{\citenamefont {Schweitzer}\ \emph
  {et~al.}(2008{\natexlab{b}})\citenamefont {Schweitzer}, \citenamefont
  {Bailey}, \citenamefont {Fischer}, \citenamefont {LeRoy}, \citenamefont
  {Lonsdorf}, \citenamefont {Whitham},\ and\ \citenamefont
  {Hart}}]{schweitzerPlantsoilmicroorganismInteractionsHeritable2008}%
  \BibitemOpen
  \bibfield  {author} {\bibinfo {author} {\bibfnamefont {J.~A.}\ \bibnamefont
  {Schweitzer}}, \bibinfo {author} {\bibfnamefont {J.~K.}\ \bibnamefont
  {Bailey}}, \bibinfo {author} {\bibfnamefont {D.~G.}\ \bibnamefont {Fischer}},
  \bibinfo {author} {\bibfnamefont {C.~J.}\ \bibnamefont {LeRoy}}, \bibinfo
  {author} {\bibfnamefont {E.~V.}\ \bibnamefont {Lonsdorf}}, \bibinfo {author}
  {\bibfnamefont {T.~G.}\ \bibnamefont {Whitham}}, \ and\ \bibinfo {author}
  {\bibfnamefont {S.~C.}\ \bibnamefont {Hart}},\ }\href {\doibase
  10.1890/07-0337.1} {\bibfield  {journal} {\bibinfo  {journal} {Ecology}\
  }\textbf {\bibinfo {volume} {89}},\ \bibinfo {pages} {773} (\bibinfo {year}
  {2008}{\natexlab{b}})}\BibitemShut {NoStop}%
\bibitem [{\citenamefont {Palkovacs}\ and\ \citenamefont
  {Hendry}(2010)}]{palkovacsEcoevolutionaryDynamicsIntertwining2010}%
  \BibitemOpen
  \bibfield  {author} {\bibinfo {author} {\bibfnamefont {E.~P.}\ \bibnamefont
  {Palkovacs}}\ and\ \bibinfo {author} {\bibfnamefont {A.~P.}\ \bibnamefont
  {Hendry}},\ }\href {\doibase 10.3410/B2-1} {\bibfield  {journal} {\bibinfo
  {journal} {F1000 Biology Reports}\ }\textbf {\bibinfo {volume} {2}} (\bibinfo
  {year} {2010}),\ 10.3410/B2-1}\BibitemShut {NoStop}%
\bibitem [{\citenamefont {Hendry}(2019)}]{hendryCritiqueEcoEvolutionary2019}%
  \BibitemOpen
  \bibfield  {author} {\bibinfo {author} {\bibfnamefont {A.~P.}\ \bibnamefont
  {Hendry}},\ }\href {\doibase 10.1111/1365-2435.13244} {\bibfield  {journal}
  {\bibinfo  {journal} {Functional Ecology}\ }\textbf {\bibinfo {volume}
  {33}},\ \bibinfo {pages} {84} (\bibinfo {year} {2019})}\BibitemShut {NoStop}%
\bibitem [{\citenamefont {Bailey}\ \emph {et~al.}(2006)\citenamefont {Bailey},
  \citenamefont {Wooley}, \citenamefont {Lindroth},\ and\ \citenamefont
  {Whitham}}]{bailey2006importance}%
  \BibitemOpen
  \bibfield  {author} {\bibinfo {author} {\bibfnamefont {J.~K.}\ \bibnamefont
  {Bailey}}, \bibinfo {author} {\bibfnamefont {S.~C.}\ \bibnamefont {Wooley}},
  \bibinfo {author} {\bibfnamefont {R.~L.}\ \bibnamefont {Lindroth}}, \ and\
  \bibinfo {author} {\bibfnamefont {T.~G.}\ \bibnamefont {Whitham}},\
  }\href@noop {} {\bibfield  {journal} {\bibinfo  {journal} {Ecology letters}\
  }\textbf {\bibinfo {volume} {9}},\ \bibinfo {pages} {78} (\bibinfo {year}
  {2006})}\BibitemShut {NoStop}%
\bibitem [{\citenamefont {Doebeli}\ and\ \citenamefont
  {Ispolatov}(2017)}]{doebeliDiversityCoevolutionaryDynamics2017}%
  \BibitemOpen
  \bibfield  {author} {\bibinfo {author} {\bibfnamefont {M.}~\bibnamefont
  {Doebeli}}\ and\ \bibinfo {author} {\bibfnamefont {I.}~\bibnamefont
  {Ispolatov}},\ }\href {\doibase 10.1086/689891} {\bibfield  {journal}
  {\bibinfo  {journal} {The American Naturalist}\ }\textbf {\bibinfo {volume}
  {189}},\ \bibinfo {pages} {105} (\bibinfo {year} {2017})}\BibitemShut
  {NoStop}%
\bibitem [{\citenamefont {Dieckmann}\ and\ \citenamefont
  {Law}()}]{dieckmannDynamicalTheoryCoevolution}%
  \BibitemOpen
  \bibfield  {author} {\bibinfo {author} {\bibfnamefont {U.}~\bibnamefont
  {Dieckmann}}\ and\ \bibinfo {author} {\bibfnamefont {R.}~\bibnamefont
  {Law}},\ }\href@noop {} {\ ,\ \bibinfo {pages} {34}}\BibitemShut {NoStop}%
\bibitem [{\citenamefont {Rogers}\ \emph
  {et~al.}(2012{\natexlab{a}})\citenamefont {Rogers}, \citenamefont {McKane},\
  and\ \citenamefont {Rossberg}}]{rogersDemographicNoiseCan2012}%
  \BibitemOpen
  \bibfield  {author} {\bibinfo {author} {\bibfnamefont {T.}~\bibnamefont
  {Rogers}}, \bibinfo {author} {\bibfnamefont {A.~J.}\ \bibnamefont {McKane}},
  \ and\ \bibinfo {author} {\bibfnamefont {A.~G.}\ \bibnamefont {Rossberg}},\
  }\href {\doibase 10.1209/0295-5075/97/40008} {\bibfield  {journal} {\bibinfo
  {journal} {EPL (Europhysics Letters)}\ }\textbf {\bibinfo {volume} {97}},\
  \bibinfo {pages} {40008} (\bibinfo {year} {2012}{\natexlab{a}})}\BibitemShut
  {NoStop}%
\bibitem [{\citenamefont {Rogers}\ and\ \citenamefont
  {McKane}(2015)}]{rogersModesCompetitionFitness2015}%
  \BibitemOpen
  \bibfield  {author} {\bibinfo {author} {\bibfnamefont {T.}~\bibnamefont
  {Rogers}}\ and\ \bibinfo {author} {\bibfnamefont {A.~J.}\ \bibnamefont
  {McKane}},\ }\href {\doibase 10.1103/PhysRevE.92.032708} {\bibfield
  {journal} {\bibinfo  {journal} {Physical Review E}\ }\textbf {\bibinfo
  {volume} {92}} (\bibinfo {year} {2015}),\
  10.1103/PhysRevE.92.032708}\BibitemShut {NoStop}%
\bibitem [{\citenamefont {Rogers}\ \emph
  {et~al.}(2012{\natexlab{b}})\citenamefont {Rogers}, \citenamefont {McKane},\
  and\ \citenamefont {Rossberg}}]{rogersSpontaneousGeneticClustering2012}%
  \BibitemOpen
  \bibfield  {author} {\bibinfo {author} {\bibfnamefont {T.}~\bibnamefont
  {Rogers}}, \bibinfo {author} {\bibfnamefont {A.~J.}\ \bibnamefont {McKane}},
  \ and\ \bibinfo {author} {\bibfnamefont {A.~G.}\ \bibnamefont {Rossberg}},\
  }\href {\doibase 10.1088/1478-3975/9/6/066002} {\bibfield  {journal}
  {\bibinfo  {journal} {Physical Biology}\ }\textbf {\bibinfo {volume} {9}},\
  \bibinfo {pages} {066002} (\bibinfo {year} {2012}{\natexlab{b}})}\BibitemShut
  {NoStop}%
\bibitem [{\citenamefont {Shnerb}(2004)}]{shnerbPatternFormationNonlocal2004}%
  \BibitemOpen
  \bibfield  {author} {\bibinfo {author} {\bibfnamefont {N.~M.}\ \bibnamefont
  {Shnerb}},\ }\href {\doibase 10.1103/PhysRevE.69.061917} {\bibfield
  {journal} {\bibinfo  {journal} {Physical Review E}\ }\textbf {\bibinfo
  {volume} {69}} (\bibinfo {year} {2004}),\
  10.1103/PhysRevE.69.061917}\BibitemShut {NoStop}%
\bibitem [{\citenamefont {Pigolotti}\ \emph {et~al.}(2007)\citenamefont
  {Pigolotti}, \citenamefont {López},\ and\ \citenamefont
  {Hernández-García}}]{pigolottiSpeciesClusteringCompetitive2007}%
  \BibitemOpen
  \bibfield  {author} {\bibinfo {author} {\bibfnamefont {S.}~\bibnamefont
  {Pigolotti}}, \bibinfo {author} {\bibfnamefont {C.}~\bibnamefont {López}}, \
  and\ \bibinfo {author} {\bibfnamefont {E.}~\bibnamefont
  {Hernández-García}},\ }\href {\doibase 10.1103/PhysRevLett.98.258101}
  {\bibfield  {journal} {\bibinfo  {journal} {Physical Review Letters}\
  }\textbf {\bibinfo {volume} {98}},\ \bibinfo {pages} {258101} (\bibinfo
  {year} {2007})}\BibitemShut {NoStop}%
\bibitem [{\citenamefont {Pigolotti}\ \emph {et~al.}(2010)\citenamefont
  {Pigolotti}, \citenamefont {López}, \citenamefont {Hernández-García},\
  and\ \citenamefont {Andersen}}]{pigolottiHowGaussianCompetition2010}%
  \BibitemOpen
  \bibfield  {author} {\bibinfo {author} {\bibfnamefont {S.}~\bibnamefont
  {Pigolotti}}, \bibinfo {author} {\bibfnamefont {C.}~\bibnamefont {López}},
  \bibinfo {author} {\bibfnamefont {E.}~\bibnamefont {Hernández-García}}, \
  and\ \bibinfo {author} {\bibfnamefont {K.~H.}\ \bibnamefont {Andersen}},\
  }\href {\doibase 10.1007/s12080-009-0056-2} {\bibfield  {journal} {\bibinfo
  {journal} {Theoretical Ecology}\ }\textbf {\bibinfo {volume} {3}},\ \bibinfo
  {pages} {89} (\bibinfo {year} {2010})}\BibitemShut {NoStop}%
\bibitem [{\citenamefont {Hernández-García}\ \emph
  {et~al.}(2009)\citenamefont {Hernández-García}, \citenamefont {López},
  \citenamefont {Pigolotti},\ and\ \citenamefont
  {Andersen}}]{hernandez-garciaSpeciesCompetitionCoexistence2009}%
  \BibitemOpen
  \bibfield  {author} {\bibinfo {author} {\bibfnamefont {E.}~\bibnamefont
  {Hernández-García}}, \bibinfo {author} {\bibfnamefont {C.}~\bibnamefont
  {López}}, \bibinfo {author} {\bibfnamefont {S.}~\bibnamefont {Pigolotti}}, \
  and\ \bibinfo {author} {\bibfnamefont {K.~H.}\ \bibnamefont {Andersen}},\
  }\href {\doibase 10.1098/rsta.2009.0086} {\bibfield  {journal} {\bibinfo
  {journal} {Philosophical Transactions of the Royal Society A: Mathematical,
  Physical and Engineering Sciences}\ }\textbf {\bibinfo {volume} {367}},\
  \bibinfo {pages} {3183} (\bibinfo {year} {2009})}\BibitemShut {NoStop}%
\bibitem [{\citenamefont {López}\ and\ \citenamefont
  {Hernández-García}(2004)}]{lopezFluctuationsImpactPatternforming2004a}%
  \BibitemOpen
  \bibfield  {author} {\bibinfo {author} {\bibfnamefont {C.}~\bibnamefont
  {López}}\ and\ \bibinfo {author} {\bibfnamefont {E.}~\bibnamefont
  {Hernández-García}},\ }\href {\doibase 10.1016/j.physd.2004.08.016}
  {\bibfield  {journal} {\bibinfo  {journal} {Physica D: Nonlinear Phenomena}\
  }\textbf {\bibinfo {volume} {199}},\ \bibinfo {pages} {223} (\bibinfo {year}
  {2004})}\BibitemShut {NoStop}%
\bibitem [{\citenamefont {Fort}\ \emph {et~al.}(2010)\citenamefont {Fort},
  \citenamefont {Scheffer},\ and\ \citenamefont
  {Van~Nes}}]{fortClumpingTransitionNiche2010}%
  \BibitemOpen
  \bibfield  {author} {\bibinfo {author} {\bibfnamefont {H.}~\bibnamefont
  {Fort}}, \bibinfo {author} {\bibfnamefont {M.}~\bibnamefont {Scheffer}}, \
  and\ \bibinfo {author} {\bibfnamefont {E.}~\bibnamefont {Van~Nes}},\ }\href
  {\doibase 10.1088/1742-5468/2010/05/P05005} {\bibfield  {journal} {\bibinfo
  {journal} {Journal of Statistical Mechanics: Theory and Experiment}\ }\textbf
  {\bibinfo {volume} {2010}},\ \bibinfo {pages} {P05005} (\bibinfo {year}
  {2010})}\BibitemShut {NoStop}%
\bibitem [{\citenamefont {Biancalani}\ \emph {et~al.}(2015)\citenamefont
  {Biancalani}, \citenamefont {DeVille},\ and\ \citenamefont
  {Goldenfeld}}]{biancalaniFrameworkAnalyzingEcological2015}%
  \BibitemOpen
  \bibfield  {author} {\bibinfo {author} {\bibfnamefont {T.}~\bibnamefont
  {Biancalani}}, \bibinfo {author} {\bibfnamefont {L.}~\bibnamefont {DeVille}},
  \ and\ \bibinfo {author} {\bibfnamefont {N.}~\bibnamefont {Goldenfeld}},\
  }\href {\doibase 10.1103/PhysRevE.91.052107} {\bibfield  {journal} {\bibinfo
  {journal} {Physical Review E}\ }\textbf {\bibinfo {volume} {91}} (\bibinfo
  {year} {2015}),\ 10.1103/PhysRevE.91.052107}\BibitemShut {NoStop}%
\bibitem [{\citenamefont {Biancalani}\ \emph {et~al.}(2017)\citenamefont
  {Biancalani}, \citenamefont {Jafarpour},\ and\ \citenamefont
  {Goldenfeld}}]{biancalaniGiantAmplificationNoise2017}%
  \BibitemOpen
  \bibfield  {author} {\bibinfo {author} {\bibfnamefont {T.}~\bibnamefont
  {Biancalani}}, \bibinfo {author} {\bibfnamefont {F.}~\bibnamefont
  {Jafarpour}}, \ and\ \bibinfo {author} {\bibfnamefont {N.}~\bibnamefont
  {Goldenfeld}},\ }\href {\doibase 10.1103/PhysRevLett.118.018101} {\bibfield
  {journal} {\bibinfo  {journal} {Physical Review Letters}\ }\textbf {\bibinfo
  {volume} {118}} (\bibinfo {year} {2017}),\
  10.1103/PhysRevLett.118.018101}\BibitemShut {NoStop}%
\bibitem [{\citenamefont {Butler}\ and\ \citenamefont
  {Goldenfeld}(2011)}]{butlerFluctuationdrivenTuringPatterns2011a}%
  \BibitemOpen
  \bibfield  {author} {\bibinfo {author} {\bibfnamefont {T.}~\bibnamefont
  {Butler}}\ and\ \bibinfo {author} {\bibfnamefont {N.}~\bibnamefont
  {Goldenfeld}},\ }\href {\doibase 10.1103/PhysRevE.84.011112} {\bibfield
  {journal} {\bibinfo  {journal} {Physical Review E}\ }\textbf {\bibinfo
  {volume} {84}} (\bibinfo {year} {2011}),\
  10.1103/PhysRevE.84.011112}\BibitemShut {NoStop}%
\bibitem [{\citenamefont {Shih}\ and\ \citenamefont
  {Goldenfeld}(2014)}]{shihPathintegralCalculationEmergence2014}%
  \BibitemOpen
  \bibfield  {author} {\bibinfo {author} {\bibfnamefont {H.-Y.}\ \bibnamefont
  {Shih}}\ and\ \bibinfo {author} {\bibfnamefont {N.}~\bibnamefont
  {Goldenfeld}},\ }\href {\doibase 10.1103/PhysRevE.90.050702} {\bibfield
  {journal} {\bibinfo  {journal} {Physical Review E}\ }\textbf {\bibinfo
  {volume} {90}} (\bibinfo {year} {2014}),\
  10.1103/PhysRevE.90.050702}\BibitemShut {NoStop}%
\bibitem [{\citenamefont
  {MacArthur}(1970{\natexlab{b}})}]{macarthur1970species}%
  \BibitemOpen
  \bibfield  {author} {\bibinfo {author} {\bibfnamefont {R.}~\bibnamefont
  {MacArthur}},\ }\href@noop {} {\bibfield  {journal} {\bibinfo  {journal}
  {Theoretical population biology}\ }\textbf {\bibinfo {volume} {1}},\ \bibinfo
  {pages} {1} (\bibinfo {year} {1970}{\natexlab{b}})}\BibitemShut {NoStop}%
\bibitem [{\citenamefont {Agranov}\ \emph
  {et~al.}(2022{\natexlab{a}})\citenamefont {Agranov}, \citenamefont {Cates},\
  and\ \citenamefont {Jack}}]{agranovEntropyProductionIts2022a}%
  \BibitemOpen
  \bibfield  {author} {\bibinfo {author} {\bibfnamefont {T.}~\bibnamefont
  {Agranov}}, \bibinfo {author} {\bibfnamefont {M.~E.}\ \bibnamefont {Cates}},
  \ and\ \bibinfo {author} {\bibfnamefont {R.~L.}\ \bibnamefont {Jack}},\
  }\href {\doibase 10.1088/1742-5468/aca0eb} {\bibfield  {journal} {\bibinfo
  {journal} {Journal of Statistical Mechanics: Theory and Experiment}\ }\textbf
  {\bibinfo {volume} {2022}},\ \bibinfo {pages} {123201} (\bibinfo {year}
  {2022}{\natexlab{a}})}\BibitemShut {NoStop}%
\bibitem [{\citenamefont {Agranov}\ \emph {et~al.}(2021)\citenamefont
  {Agranov}, \citenamefont {Ro}, \citenamefont {Kafri},\ and\ \citenamefont
  {Lecomte}}]{agranovExactFluctuatingHydrodynamics2021}%
  \BibitemOpen
  \bibfield  {author} {\bibinfo {author} {\bibfnamefont {T.}~\bibnamefont
  {Agranov}}, \bibinfo {author} {\bibfnamefont {S.}~\bibnamefont {Ro}},
  \bibinfo {author} {\bibfnamefont {Y.}~\bibnamefont {Kafri}}, \ and\ \bibinfo
  {author} {\bibfnamefont {V.}~\bibnamefont {Lecomte}},\ }\href {\doibase
  10.1088/1742-5468/ac1406} {\bibfield  {journal} {\bibinfo  {journal} {Journal
  of Statistical Mechanics: Theory and Experiment}\ }\textbf {\bibinfo {volume}
  {2021}},\ \bibinfo {pages} {083208} (\bibinfo {year} {2021})}\BibitemShut
  {NoStop}%
\bibitem [{\citenamefont {Agranov}\ \emph
  {et~al.}(2022{\natexlab{b}})\citenamefont {Agranov}, \citenamefont {Ro},
  \citenamefont {Kafri},\ and\ \citenamefont
  {Lecomte}}]{agranovMacroscopicFluctuationTheory2022}%
  \BibitemOpen
  \bibfield  {author} {\bibinfo {author} {\bibfnamefont {T.}~\bibnamefont
  {Agranov}}, \bibinfo {author} {\bibfnamefont {S.}~\bibnamefont {Ro}},
  \bibinfo {author} {\bibfnamefont {Y.}~\bibnamefont {Kafri}}, \ and\ \bibinfo
  {author} {\bibfnamefont {V.}~\bibnamefont {Lecomte}},\ }\href {\doibase
  10.48550/ARXIV.2208.02124} {\  (\bibinfo {year} {2022}{\natexlab{b}}),\
  10.48550/ARXIV.2208.02124},\ \bibinfo {note} {publisher: arXiv Version
  Number: 2}\BibitemShut {NoStop}%
\bibitem [{\citenamefont {Agranov}\ \emph {et~al.}(2023)\citenamefont
  {Agranov}, \citenamefont {Cates},\ and\ \citenamefont
  {Jack}}]{agranovTricriticalBehaviorDynamical2023}%
  \BibitemOpen
  \bibfield  {author} {\bibinfo {author} {\bibfnamefont {T.}~\bibnamefont
  {Agranov}}, \bibinfo {author} {\bibfnamefont {M.~E.}\ \bibnamefont {Cates}},
  \ and\ \bibinfo {author} {\bibfnamefont {R.~L.}\ \bibnamefont {Jack}},\
  }\href {\doibase 10.1103/PhysRevLett.131.017102} {\bibfield  {journal}
  {\bibinfo  {journal} {Physical Review Letters}\ }\textbf {\bibinfo {volume}
  {131}},\ \bibinfo {pages} {017102} (\bibinfo {year} {2023})}\BibitemShut
  {NoStop}%
\bibitem [{\citenamefont {Baek}\ \emph {et~al.}(2018)\citenamefont {Baek},
  \citenamefont {Kafri},\ and\ \citenamefont
  {Lecomte}}]{baekDynamicalPhaseTransitions2018}%
  \BibitemOpen
  \bibfield  {author} {\bibinfo {author} {\bibfnamefont {Y.}~\bibnamefont
  {Baek}}, \bibinfo {author} {\bibfnamefont {Y.}~\bibnamefont {Kafri}}, \ and\
  \bibinfo {author} {\bibfnamefont {V.}~\bibnamefont {Lecomte}},\ }\href
  {\doibase 10.1088/1751-8121/aaa8f9} {\bibfield  {journal} {\bibinfo
  {journal} {Journal of Physics A: Mathematical and Theoretical}\ }\textbf
  {\bibinfo {volume} {51}},\ \bibinfo {pages} {105001} (\bibinfo {year}
  {2018})}\BibitemShut {NoStop}%
\bibitem [{\citenamefont {Baek}\ \emph {et~al.}(2017)\citenamefont {Baek},
  \citenamefont {Kafri},\ and\ \citenamefont
  {Lecomte}}]{baekDynamicalSymmetryBreaking2017}%
  \BibitemOpen
  \bibfield  {author} {\bibinfo {author} {\bibfnamefont {Y.}~\bibnamefont
  {Baek}}, \bibinfo {author} {\bibfnamefont {Y.}~\bibnamefont {Kafri}}, \ and\
  \bibinfo {author} {\bibfnamefont {V.}~\bibnamefont {Lecomte}},\ }\href
  {\doibase 10.1103/PhysRevLett.118.030604} {\bibfield  {journal} {\bibinfo
  {journal} {Physical Review Letters}\ }\textbf {\bibinfo {volume} {118}},\
  \bibinfo {pages} {030604} (\bibinfo {year} {2017})}\BibitemShut {NoStop}%
\bibitem [{\citenamefont {Bodineau}\ and\ \citenamefont
  {Derrida}(2006)}]{bodineauCurrentLargeDeviations2006}%
  \BibitemOpen
  \bibfield  {author} {\bibinfo {author} {\bibfnamefont {T.}~\bibnamefont
  {Bodineau}}\ and\ \bibinfo {author} {\bibfnamefont {B.}~\bibnamefont
  {Derrida}},\ }\href {\doibase 10.1007/s10955-006-9048-4} {\bibfield
  {journal} {\bibinfo  {journal} {Journal of Statistical Physics}\ }\textbf
  {\bibinfo {volume} {123}},\ \bibinfo {pages} {277} (\bibinfo {year}
  {2006})}\BibitemShut {NoStop}%
\bibitem [{\citenamefont {Bodineau}\ and\ \citenamefont
  {Derrida}(2005)}]{bodineauDistributionCurrentNonequilibrium2005}%
  \BibitemOpen
  \bibfield  {author} {\bibinfo {author} {\bibfnamefont {T.}~\bibnamefont
  {Bodineau}}\ and\ \bibinfo {author} {\bibfnamefont {B.}~\bibnamefont
  {Derrida}},\ }\href {\doibase 10.1103/PhysRevE.72.066110} {\bibfield
  {journal} {\bibinfo  {journal} {Physical Review E}\ }\textbf {\bibinfo
  {volume} {72}} (\bibinfo {year} {2005}),\
  10.1103/PhysRevE.72.066110}\BibitemShut {NoStop}%
\bibitem [{\citenamefont {Appert-Rolland}\ \emph {et~al.}(2008)\citenamefont
  {Appert-Rolland}, \citenamefont {Derrida}, \citenamefont {Lecomte},\ and\
  \citenamefont {Van~Wijland}}]{appert-rollandUniversalCumulantsCurrent2008}%
  \BibitemOpen
  \bibfield  {author} {\bibinfo {author} {\bibfnamefont {C.}~\bibnamefont
  {Appert-Rolland}}, \bibinfo {author} {\bibfnamefont {B.}~\bibnamefont
  {Derrida}}, \bibinfo {author} {\bibfnamefont {V.}~\bibnamefont {Lecomte}}, \
  and\ \bibinfo {author} {\bibfnamefont {F.}~\bibnamefont {Van~Wijland}},\
  }\href {\doibase 10.1103/PhysRevE.78.021122} {\bibfield  {journal} {\bibinfo
  {journal} {Physical Review E}\ }\textbf {\bibinfo {volume} {78}},\ \bibinfo
  {pages} {021122} (\bibinfo {year} {2008})},\ \bibinfo {note} {arXiv:
  0804.2590}\BibitemShut {NoStop}%
\bibitem [{\citenamefont {Krašovec}\ \emph {et~al.}(2017)\citenamefont
  {Krašovec}, \citenamefont {Richards}, \citenamefont {Gifford}, \citenamefont
  {Hatcher}, \citenamefont {Faulkner}, \citenamefont {Belavkin}, \citenamefont
  {Channon}, \citenamefont {Aston}, \citenamefont {McBain},\ and\ \citenamefont
  {Knight}}]{krasovecSpontaneousMutationRate2017}%
  \BibitemOpen
  \bibfield  {author} {\bibinfo {author} {\bibfnamefont {R.}~\bibnamefont
  {Krašovec}}, \bibinfo {author} {\bibfnamefont {H.}~\bibnamefont {Richards}},
  \bibinfo {author} {\bibfnamefont {D.~R.}\ \bibnamefont {Gifford}}, \bibinfo
  {author} {\bibfnamefont {C.}~\bibnamefont {Hatcher}}, \bibinfo {author}
  {\bibfnamefont {K.~J.}\ \bibnamefont {Faulkner}}, \bibinfo {author}
  {\bibfnamefont {R.~V.}\ \bibnamefont {Belavkin}}, \bibinfo {author}
  {\bibfnamefont {A.}~\bibnamefont {Channon}}, \bibinfo {author} {\bibfnamefont
  {E.}~\bibnamefont {Aston}}, \bibinfo {author} {\bibfnamefont {A.~J.}\
  \bibnamefont {McBain}}, \ and\ \bibinfo {author} {\bibfnamefont {C.~G.}\
  \bibnamefont {Knight}},\ }\href {\doibase 10.1371/journal.pbio.2002731}
  {\bibfield  {journal} {\bibinfo  {journal} {PLOS Biology}\ }\textbf {\bibinfo
  {volume} {15}},\ \bibinfo {pages} {e2002731} (\bibinfo {year}
  {2017})}\BibitemShut {NoStop}%
\bibitem [{\citenamefont {Aanen}\ and\ \citenamefont
  {Debets}(2019)}]{aanenMutationratePlasticityGermline2019}%
  \BibitemOpen
  \bibfield  {author} {\bibinfo {author} {\bibfnamefont {D.~K.}\ \bibnamefont
  {Aanen}}\ and\ \bibinfo {author} {\bibfnamefont {A.~J.~M.}\ \bibnamefont
  {Debets}},\ }\href {\doibase 10.1098/rspb.2019.0128} {\bibfield  {journal}
  {\bibinfo  {journal} {Proceedings of the Royal Society B: Biological
  Sciences}\ }\textbf {\bibinfo {volume} {286}},\ \bibinfo {pages} {20190128}
  (\bibinfo {year} {2019})}\BibitemShut {NoStop}%
\bibitem [{\citenamefont {Cross}\ and\ \citenamefont
  {Hohenberg}(1993{\natexlab{a}})}]{crossPatternFormationOutside1993}%
  \BibitemOpen
  \bibfield  {author} {\bibinfo {author} {\bibfnamefont {M.~C.}\ \bibnamefont
  {Cross}}\ and\ \bibinfo {author} {\bibfnamefont {P.~C.}\ \bibnamefont
  {Hohenberg}},\ }\href {\doibase 10.1103/RevModPhys.65.851} {\bibfield
  {journal} {\bibinfo  {journal} {Reviews of Modern Physics}\ }\textbf
  {\bibinfo {volume} {65}},\ \bibinfo {pages} {851} (\bibinfo {year}
  {1993}{\natexlab{a}})}\BibitemShut {NoStop}%
\bibitem [{\citenamefont {Jiang}\ and\ \citenamefont
  {Wang}(2019)}]{jiangTraitspacePatterningRole2019a}%
  \BibitemOpen
  \bibfield  {author} {\bibinfo {author} {\bibfnamefont {H.}~\bibnamefont
  {Jiang}}\ and\ \bibinfo {author} {\bibfnamefont {S.}~\bibnamefont {Wang}},\
  }\href {\doibase 10.1103/PhysRevResearch.1.033164} {\bibfield  {journal}
  {\bibinfo  {journal} {Physical Review Research}\ }\textbf {\bibinfo {volume}
  {1}} (\bibinfo {year} {2019}),\ 10.1103/PhysRevResearch.1.033164}\BibitemShut
  {NoStop}%
\bibitem [{\citenamefont
  {Touchette}(2009)}]{touchetteLargeDeviationApproach2009}%
  \BibitemOpen
  \bibfield  {author} {\bibinfo {author} {\bibfnamefont {H.}~\bibnamefont
  {Touchette}},\ }\href {\doibase 10.1016/j.physrep.2009.05.002} {\bibfield
  {journal} {\bibinfo  {journal} {Physics Reports}\ }\textbf {\bibinfo {volume}
  {478}},\ \bibinfo {pages} {1} (\bibinfo {year} {2009})}\BibitemShut {NoStop}%
\bibitem [{\citenamefont {Touchette}\ and\ \citenamefont
  {Harris}(2013)}]{touchetteLargeDeviationApproach2013}%
  \BibitemOpen
  \bibfield  {author} {\bibinfo {author} {\bibfnamefont {H.}~\bibnamefont
  {Touchette}}\ and\ \bibinfo {author} {\bibfnamefont {R.~J.}\ \bibnamefont
  {Harris}},\ }in\ \href {\doibase 10.1002/9783527658701.ch11}
  {{\selectlanguage {english}\emph {\bibinfo {booktitle} {Nonequilibrium
  {Statistical} {Physics} of {Small} {Systems}}}}},\ \bibinfo {editor} {edited
  by\ \bibinfo {editor} {\bibfnamefont {R.}~\bibnamefont {Klages}}, \bibinfo
  {editor} {\bibfnamefont {W.}~\bibnamefont {Just}}, \ and\ \bibinfo {editor}
  {\bibfnamefont {C.}~\bibnamefont {Jarzynski}}}\ (\bibinfo  {publisher}
  {Wiley-VCH Verlag GmbH \& Co. KGaA},\ \bibinfo {address} {Weinheim,
  Germany},\ \bibinfo {year} {2013})\ pp.\ \bibinfo {pages}
  {335--360}\BibitemShut {NoStop}%
\bibitem [{\citenamefont {Martin}\ \emph {et~al.}(1973)\citenamefont {Martin},
  \citenamefont {Siggia},\ and\ \citenamefont
  {Rose}}]{martinStatisticalDynamicsClassical1973}%
  \BibitemOpen
  \bibfield  {author} {\bibinfo {author} {\bibfnamefont {P.~C.}\ \bibnamefont
  {Martin}}, \bibinfo {author} {\bibfnamefont {E.~D.}\ \bibnamefont {Siggia}},
  \ and\ \bibinfo {author} {\bibfnamefont {H.~A.}\ \bibnamefont {Rose}},\
  }\href {\doibase 10.1103/PhysRevA.8.423} {\bibfield  {journal} {\bibinfo
  {journal} {Physical Review A}\ }\textbf {\bibinfo {volume} {8}},\ \bibinfo
  {pages} {423} (\bibinfo {year} {1973})}\BibitemShut {NoStop}%
\bibitem [{\citenamefont
  {Janssen}(1976)}]{janssenLagrangeanClassicalField1976}%
  \BibitemOpen
  \bibfield  {author} {\bibinfo {author} {\bibfnamefont {H.-K.}\ \bibnamefont
  {Janssen}},\ }\href {\doibase 10.1007/BF01316547} {\bibfield  {journal}
  {\bibinfo  {journal} {Zeitschrift Physik B Condensed Matter and
  Quanta}\ }\textbf {\bibinfo {volume} {23}},\ \bibinfo {pages} {377} (\bibinfo
  {year} {1976})}\BibitemShut {NoStop}%
\bibitem [{\citenamefont
  {De~Dominicis}(1976)}]{dedominicisTechniquesRenormalisationTheorie1976}%
  \BibitemOpen
  \bibfield  {author} {\bibinfo {author} {\bibfnamefont {C.}~\bibnamefont
  {De~Dominicis}},\ }\href {\doibase 10.1051/jphyscol:1976138} {\bibfield
  {journal} {\bibinfo  {journal} {Le Journal de Physique Colloques}\ }\textbf
  {\bibinfo {volume} {37}},\ \bibinfo {pages} {C1} (\bibinfo {year}
  {1976})}\BibitemShut {NoStop}%
\bibitem [{\citenamefont {De~Dominicis}\ and\ \citenamefont
  {Peliti}(1978)}]{dedominicisFieldtheoryRenormalizationCritical1978}%
  \BibitemOpen
  \bibfield  {author} {\bibinfo {author} {\bibfnamefont {C.}~\bibnamefont
  {De~Dominicis}}\ and\ \bibinfo {author} {\bibfnamefont {L.}~\bibnamefont
  {Peliti}},\ }\href {\doibase 10.1103/PhysRevB.18.353} {\bibfield  {journal}
  {\bibinfo  {journal} {Physical Review B}\ }\textbf {\bibinfo {volume} {18}},\
  \bibinfo {pages} {353} (\bibinfo {year} {1978})}\BibitemShut {NoStop}%
\bibitem [{\citenamefont {Bertini}\ \emph {et~al.}(2005)\citenamefont
  {Bertini}, \citenamefont {De~Sole}, \citenamefont {Gabrielli}, \citenamefont
  {Jona-Lasinio},\ and\ \citenamefont
  {Landim}}]{bertiniCurrentFluctuationsStochastic2005}%
  \BibitemOpen
  \bibfield  {author} {\bibinfo {author} {\bibfnamefont {L.}~\bibnamefont
  {Bertini}}, \bibinfo {author} {\bibfnamefont {A.}~\bibnamefont {De~Sole}},
  \bibinfo {author} {\bibfnamefont {D.}~\bibnamefont {Gabrielli}}, \bibinfo
  {author} {\bibfnamefont {G.}~\bibnamefont {Jona-Lasinio}}, \ and\ \bibinfo
  {author} {\bibfnamefont {C.}~\bibnamefont {Landim}},\ }\href {\doibase
  10.1103/PhysRevLett.94.030601} {\bibfield  {journal} {\bibinfo  {journal}
  {Physical Review Letters}\ }\textbf {\bibinfo {volume} {94}},\ \bibinfo
  {pages} {030601} (\bibinfo {year} {2005})}\BibitemShut {NoStop}%
\bibitem [{\citenamefont {Giardinà}\ \emph {et~al.}(2006)\citenamefont
  {Giardinà}, \citenamefont {Kurchan},\ and\ \citenamefont
  {Peliti}}]{giardinaDirectEvaluationLargeDeviation2006}%
  \BibitemOpen
  \bibfield  {author} {\bibinfo {author} {\bibfnamefont {C.}~\bibnamefont
  {Giardinà}}, \bibinfo {author} {\bibfnamefont {J.}~\bibnamefont {Kurchan}},
  \ and\ \bibinfo {author} {\bibfnamefont {L.}~\bibnamefont {Peliti}},\ }\href
  {\doibase 10.1103/PhysRevLett.96.120603} {\bibfield  {journal} {\bibinfo
  {journal} {Physical Review Letters}\ }\textbf {\bibinfo {volume} {96}},\
  \bibinfo {pages} {120603} (\bibinfo {year} {2006})}\BibitemShut {NoStop}%
\bibitem [{\citenamefont {Giardina}\ \emph {et~al.}(2011)\citenamefont
  {Giardina}, \citenamefont {Kurchan}, \citenamefont {Lecomte},\ and\
  \citenamefont {Tailleur}}]{giardinaSimulatingRareEvents2011}%
  \BibitemOpen
  \bibfield  {author} {\bibinfo {author} {\bibfnamefont {C.}~\bibnamefont
  {Giardina}}, \bibinfo {author} {\bibfnamefont {J.}~\bibnamefont {Kurchan}},
  \bibinfo {author} {\bibfnamefont {V.}~\bibnamefont {Lecomte}}, \ and\
  \bibinfo {author} {\bibfnamefont {J.}~\bibnamefont {Tailleur}},\ }\href
  {\doibase 10.1007/s10955-011-0350-4} {\bibfield  {journal} {\bibinfo
  {journal} {Journal of Statistical Physics}\ }\textbf {\bibinfo {volume}
  {145}},\ \bibinfo {pages} {787} (\bibinfo {year} {2011})}\BibitemShut
  {NoStop}%
\bibitem [{\citenamefont {Tailleur}\ and\ \citenamefont
  {Kurchan}(2007)}]{tailleurProbingRarePhysical2007}%
  \BibitemOpen
  \bibfield  {author} {\bibinfo {author} {\bibfnamefont {J.}~\bibnamefont
  {Tailleur}}\ and\ \bibinfo {author} {\bibfnamefont {J.}~\bibnamefont
  {Kurchan}},\ }\href {\doibase 10.1038/nphys515} {\bibfield  {journal}
  {\bibinfo  {journal} {Nature Physics}\ }\textbf {\bibinfo {volume} {3}},\
  \bibinfo {pages} {203} (\bibinfo {year} {2007})}\BibitemShut {NoStop}%
\bibitem [{\citenamefont {Nemoto}\ \emph {et~al.}(2017)\citenamefont {Nemoto},
  \citenamefont {Jack},\ and\ \citenamefont
  {Lecomte}}]{nemotoFiniteSizeScalingFirstOrder2017}%
  \BibitemOpen
  \bibfield  {author} {\bibinfo {author} {\bibfnamefont {T.}~\bibnamefont
  {Nemoto}}, \bibinfo {author} {\bibfnamefont {R.~L.}\ \bibnamefont {Jack}}, \
  and\ \bibinfo {author} {\bibfnamefont {V.}~\bibnamefont {Lecomte}},\ }\href
  {\doibase 10.1103/PhysRevLett.118.115702} {\bibfield  {journal} {\bibinfo
  {journal} {Physical Review Letters}\ }\textbf {\bibinfo {volume} {118}},\
  \bibinfo {pages} {115702} (\bibinfo {year} {2017})}\BibitemShut {NoStop}%
\bibitem [{\citenamefont {Nemoto}\ \emph {et~al.}(2016)\citenamefont {Nemoto},
  \citenamefont {Bouchet}, \citenamefont {Jack},\ and\ \citenamefont
  {Lecomte}}]{nemotoPopulationdynamicsMethodMulticanonical2016}%
  \BibitemOpen
  \bibfield  {author} {\bibinfo {author} {\bibfnamefont {T.}~\bibnamefont
  {Nemoto}}, \bibinfo {author} {\bibfnamefont {F.}~\bibnamefont {Bouchet}},
  \bibinfo {author} {\bibfnamefont {R.~L.}\ \bibnamefont {Jack}}, \ and\
  \bibinfo {author} {\bibfnamefont {V.}~\bibnamefont {Lecomte}},\ }\href
  {\doibase 10.1103/PhysRevE.93.062123} {\bibfield  {journal} {\bibinfo
  {journal} {Physical Review E}\ }\textbf {\bibinfo {volume} {93}},\ \bibinfo
  {pages} {062123} (\bibinfo {year} {2016})}\BibitemShut {NoStop}%
\bibitem [{\citenamefont {Jack}\ \emph {et~al.}(2015)\citenamefont {Jack},
  \citenamefont {Thompson},\ and\ \citenamefont
  {Sollich}}]{jackHyperuniformityPhaseSeparation2015}%
  \BibitemOpen
  \bibfield  {author} {\bibinfo {author} {\bibfnamefont {R.~L.}\ \bibnamefont
  {Jack}}, \bibinfo {author} {\bibfnamefont {I.~R.}\ \bibnamefont {Thompson}},
  \ and\ \bibinfo {author} {\bibfnamefont {P.}~\bibnamefont {Sollich}},\ }\href
  {\doibase 10.1103/PhysRevLett.114.060601} {\bibfield  {journal} {\bibinfo
  {journal} {Physical Review Letters}\ }\textbf {\bibinfo {volume} {114}},\
  \bibinfo {pages} {060601} (\bibinfo {year} {2015})},\ \bibinfo {note} {arXiv:
  1409.3986}\BibitemShut {NoStop}%
\bibitem [{\citenamefont {Dolezal}\ and\ \citenamefont
  {Jack}(2019{\natexlab{a}})}]{dolezalLargeDeviationsOptimal2019}%
  \BibitemOpen
  \bibfield  {author} {\bibinfo {author} {\bibfnamefont {J.}~\bibnamefont
  {Dolezal}}\ and\ \bibinfo {author} {\bibfnamefont {R.~L.}\ \bibnamefont
  {Jack}},\ }\href {\doibase 10.1088/1742-5468/ab4801} {\bibfield  {journal}
  {\bibinfo  {journal} {Journal of Statistical Mechanics: Theory and
  Experiment}\ }\textbf {\bibinfo {volume} {2019}},\ \bibinfo {pages} {123208}
  (\bibinfo {year} {2019}{\natexlab{a}})}\BibitemShut {NoStop}%
\bibitem [{\citenamefont {GrandPre}\ \emph {et~al.}(2021)\citenamefont
  {GrandPre}, \citenamefont {Klymko}, \citenamefont {Mandadapu},\ and\
  \citenamefont {Limmer}}]{grandpreEntropyProductionFluctuations2021}%
  \BibitemOpen
  \bibfield  {author} {\bibinfo {author} {\bibfnamefont {T.}~\bibnamefont
  {GrandPre}}, \bibinfo {author} {\bibfnamefont {K.}~\bibnamefont {Klymko}},
  \bibinfo {author} {\bibfnamefont {K.~K.}\ \bibnamefont {Mandadapu}}, \ and\
  \bibinfo {author} {\bibfnamefont {D.~T.}\ \bibnamefont {Limmer}},\ }\href
  {\doibase 10.1103/PhysRevE.103.012613} {\bibfield  {journal} {\bibinfo
  {journal} {Physical Review E}\ }\textbf {\bibinfo {volume} {103}},\ \bibinfo
  {pages} {012613} (\bibinfo {year} {2021})},\ \bibinfo {note} {arXiv:
  2007.12149}\BibitemShut {NoStop}%
\bibitem [{\citenamefont {Derrida}(2007)}]{derridaNonEquilibriumSteady2007}%
  \BibitemOpen
  \bibfield  {author} {\bibinfo {author} {\bibfnamefont {B.}~\bibnamefont
  {Derrida}},\ }\href {\doibase 10.1088/1742-5468/2007/07/P07023} {\bibfield
  {journal} {\bibinfo  {journal} {Journal of Statistical Mechanics: Theory and
  Experiment}\ }\textbf {\bibinfo {volume} {2007}},\ \bibinfo {pages} {P07023}
  (\bibinfo {year} {2007})},\ \bibinfo {note} {arXiv:
  cond-mat/0703762}\BibitemShut {NoStop}%
\bibitem [{\citenamefont {Brewer}\ \emph {et~al.}(2018)\citenamefont {Brewer},
  \citenamefont {Clark}, \citenamefont {Bradford},\ and\ \citenamefont
  {Jack}}]{brewerEfficientCharacterisationLarge2018}%
  \BibitemOpen
  \bibfield  {author} {\bibinfo {author} {\bibfnamefont {T.}~\bibnamefont
  {Brewer}}, \bibinfo {author} {\bibfnamefont {S.~R.}\ \bibnamefont {Clark}},
  \bibinfo {author} {\bibfnamefont {R.}~\bibnamefont {Bradford}}, \ and\
  \bibinfo {author} {\bibfnamefont {R.~L.}\ \bibnamefont {Jack}},\ }\href
  {\doibase 10.1088/1742-5468/aab3ef} {\bibfield  {journal} {\bibinfo
  {journal} {Journal of Statistical Mechanics: Theory and Experiment}\ }\textbf
  {\bibinfo {volume} {2018}},\ \bibinfo {pages} {053204} (\bibinfo {year}
  {2018})}\BibitemShut {NoStop}%
\bibitem [{\citenamefont {Dolezal}\ and\ \citenamefont
  {Jack}(2019{\natexlab{b}})}]{dolezal2019large}%
  \BibitemOpen
  \bibfield  {author} {\bibinfo {author} {\bibfnamefont {J.}~\bibnamefont
  {Dolezal}}\ and\ \bibinfo {author} {\bibfnamefont {R.~L.}\ \bibnamefont
  {Jack}},\ }\href@noop {} {\bibfield  {journal} {\bibinfo  {journal} {Journal
  of Statistical Mechanics: Theory and Experiment}\ }\textbf {\bibinfo {volume}
  {2019}},\ \bibinfo {pages} {123208} (\bibinfo {year}
  {2019}{\natexlab{b}})}\BibitemShut {NoStop}%
\bibitem [{\citenamefont {Yan}\ and\ \citenamefont
  {Rotskoff}(2022)}]{yan2022physics}%
  \BibitemOpen
  \bibfield  {author} {\bibinfo {author} {\bibfnamefont {J.}~\bibnamefont
  {Yan}}\ and\ \bibinfo {author} {\bibfnamefont {G.~M.}\ \bibnamefont
  {Rotskoff}},\ }\href@noop {} {\bibfield  {journal} {\bibinfo  {journal} {The
  Journal of Chemical Physics}\ }\textbf {\bibinfo {volume} {157}} (\bibinfo
  {year} {2022})}\BibitemShut {NoStop}%
\bibitem [{\citenamefont {Ray}\ \emph {et~al.}(2018)\citenamefont {Ray},
  \citenamefont {Chan},\ and\ \citenamefont {Limmer}}]{ray2018exact}%
  \BibitemOpen
  \bibfield  {author} {\bibinfo {author} {\bibfnamefont {U.}~\bibnamefont
  {Ray}}, \bibinfo {author} {\bibfnamefont {G.~K.-L.}\ \bibnamefont {Chan}}, \
  and\ \bibinfo {author} {\bibfnamefont {D.~T.}\ \bibnamefont {Limmer}},\
  }\href@noop {} {\bibfield  {journal} {\bibinfo  {journal} {Physical review
  letters}\ }\textbf {\bibinfo {volume} {120}},\ \bibinfo {pages} {210602}
  (\bibinfo {year} {2018})}\BibitemShut {NoStop}%
\bibitem [{\citenamefont {Rao}\ and\ \citenamefont
  {Leibler}(2022)}]{rao2022evolutionary}%
  \BibitemOpen
  \bibfield  {author} {\bibinfo {author} {\bibfnamefont {R.}~\bibnamefont
  {Rao}}\ and\ \bibinfo {author} {\bibfnamefont {S.}~\bibnamefont {Leibler}},\
  }\href@noop {} {\bibfield  {journal} {\bibinfo  {journal} {Proceedings of the
  National Academy of Sciences}\ }\textbf {\bibinfo {volume} {119}},\ \bibinfo
  {pages} {e2112083119} (\bibinfo {year} {2022})}\BibitemShut {NoStop}%
\bibitem [{\citenamefont {Nourmohammad}\ and\ \citenamefont
  {Eksin}(2021)}]{nourmohammad2021optimal}%
  \BibitemOpen
  \bibfield  {author} {\bibinfo {author} {\bibfnamefont {A.}~\bibnamefont
  {Nourmohammad}}\ and\ \bibinfo {author} {\bibfnamefont {C.}~\bibnamefont
  {Eksin}},\ }\href@noop {} {\bibfield  {journal} {\bibinfo  {journal}
  {Physical Review X}\ }\textbf {\bibinfo {volume} {11}},\ \bibinfo {pages}
  {011044} (\bibinfo {year} {2021})}\BibitemShut {NoStop}%
\bibitem [{\citenamefont {Saha}\ \emph {et~al.}(2020)\citenamefont {Saha},
  \citenamefont {Agudo-Canalejo},\ and\ \citenamefont
  {Golestanian}}]{sahaScalarActiveMixtures2020}%
  \BibitemOpen
  \bibfield  {author} {\bibinfo {author} {\bibfnamefont {S.}~\bibnamefont
  {Saha}}, \bibinfo {author} {\bibfnamefont {J.}~\bibnamefont
  {Agudo-Canalejo}}, \ and\ \bibinfo {author} {\bibfnamefont {R.}~\bibnamefont
  {Golestanian}},\ }\href {\doibase 10.1103/PhysRevX.10.041009} {\bibfield
  {journal} {\bibinfo  {journal} {Physical Review X}\ }\textbf {\bibinfo
  {volume} {10}},\ \bibinfo {pages} {041009} (\bibinfo {year}
  {2020})}\BibitemShut {NoStop}%
\bibitem [{\citenamefont {You}\ \emph {et~al.}(2020)\citenamefont {You},
  \citenamefont {Baskaran},\ and\ \citenamefont
  {Marchetti}}]{youNonreciprocityGenericRoute2020}%
  \BibitemOpen
  \bibfield  {author} {\bibinfo {author} {\bibfnamefont {Z.}~\bibnamefont
  {You}}, \bibinfo {author} {\bibfnamefont {A.}~\bibnamefont {Baskaran}}, \
  and\ \bibinfo {author} {\bibfnamefont {M.~C.}\ \bibnamefont {Marchetti}},\
  }\href {\doibase 10.1073/pnas.2010318117} {\bibfield  {journal} {\bibinfo
  {journal} {Proceedings of the National Academy of Sciences}\ }\textbf
  {\bibinfo {volume} {117}},\ \bibinfo {pages} {19767} (\bibinfo {year}
  {2020})}\BibitemShut {NoStop}%
\bibitem [{\citenamefont {Fruchart}\ \emph {et~al.}(2021)\citenamefont
  {Fruchart}, \citenamefont {Hanai}, \citenamefont {Littlewood},\ and\
  \citenamefont {Vitelli}}]{fruchartNonreciprocalPhaseTransitions2021}%
  \BibitemOpen
  \bibfield  {author} {\bibinfo {author} {\bibfnamefont {M.}~\bibnamefont
  {Fruchart}}, \bibinfo {author} {\bibfnamefont {R.}~\bibnamefont {Hanai}},
  \bibinfo {author} {\bibfnamefont {P.~B.}\ \bibnamefont {Littlewood}}, \ and\
  \bibinfo {author} {\bibfnamefont {V.}~\bibnamefont {Vitelli}},\ }\href
  {\doibase 10.1038/s41586-021-03375-9} {\bibfield  {journal} {\bibinfo
  {journal} {Nature}\ }\textbf {\bibinfo {volume} {592}},\ \bibinfo {pages}
  {363} (\bibinfo {year} {2021})}\BibitemShut {NoStop}%
\bibitem [{\citenamefont {El-Ganainy}\ \emph {et~al.}(2018)\citenamefont
  {El-Ganainy}, \citenamefont {Makris}, \citenamefont {Khajavikhan},
  \citenamefont {Musslimani}, \citenamefont {Rotter},\ and\ \citenamefont
  {Christodoulides}}]{el2018non}%
  \BibitemOpen
  \bibfield  {author} {\bibinfo {author} {\bibfnamefont {R.}~\bibnamefont
  {El-Ganainy}}, \bibinfo {author} {\bibfnamefont {K.~G.}\ \bibnamefont
  {Makris}}, \bibinfo {author} {\bibfnamefont {M.}~\bibnamefont {Khajavikhan}},
  \bibinfo {author} {\bibfnamefont {Z.~H.}\ \bibnamefont {Musslimani}},
  \bibinfo {author} {\bibfnamefont {S.}~\bibnamefont {Rotter}}, \ and\ \bibinfo
  {author} {\bibfnamefont {D.~N.}\ \bibnamefont {Christodoulides}},\
  }\href@noop {} {\bibfield  {journal} {\bibinfo  {journal} {Nature Physics}\
  }\textbf {\bibinfo {volume} {14}},\ \bibinfo {pages} {11} (\bibinfo {year}
  {2018})}\BibitemShut {NoStop}%
\bibitem [{\citenamefont {Krasnok}\ \emph {et~al.}(2021)\citenamefont
  {Krasnok}, \citenamefont {Nefedkin},\ and\ \citenamefont
  {Alu}}]{krasnok2021parity}%
  \BibitemOpen
  \bibfield  {author} {\bibinfo {author} {\bibfnamefont {A.}~\bibnamefont
  {Krasnok}}, \bibinfo {author} {\bibfnamefont {N.}~\bibnamefont {Nefedkin}}, \
  and\ \bibinfo {author} {\bibfnamefont {A.}~\bibnamefont {Alu}},\ }\href@noop
  {} {\bibfield  {journal} {\bibinfo  {journal} {arXiv preprint
  arXiv:2103.08135}\ } (\bibinfo {year} {2021})}\BibitemShut {NoStop}%
\bibitem [{\citenamefont {Cross}\ and\ \citenamefont
  {Hohenberg}(1993{\natexlab{b}})}]{cross1993pattern}%
  \BibitemOpen
  \bibfield  {author} {\bibinfo {author} {\bibfnamefont {M.~C.}\ \bibnamefont
  {Cross}}\ and\ \bibinfo {author} {\bibfnamefont {P.~C.}\ \bibnamefont
  {Hohenberg}},\ }\href@noop {} {\bibfield  {journal} {\bibinfo  {journal}
  {Reviews of modern physics}\ }\textbf {\bibinfo {volume} {65}},\ \bibinfo
  {pages} {851} (\bibinfo {year} {1993}{\natexlab{b}})}\BibitemShut {NoStop}%
\bibitem [{\citenamefont {Cross}\ and\ \citenamefont
  {Greenside}(2009)}]{cross2009pattern}%
  \BibitemOpen
  \bibfield  {author} {\bibinfo {author} {\bibfnamefont {M.}~\bibnamefont
  {Cross}}\ and\ \bibinfo {author} {\bibfnamefont {H.}~\bibnamefont
  {Greenside}},\ }\href@noop {} {\emph {\bibinfo {title} {Pattern formation and
  dynamics in nonequilibrium systems}}}\ (\bibinfo  {publisher} {Cambridge
  University Press},\ \bibinfo {year} {2009})\BibitemShut {NoStop}%
\bibitem [{\citenamefont {Mustonen}\ and\ \citenamefont
  {Lässig}(2009)}]{mustonenFitnessLandscapesSeascapes2009}%
  \BibitemOpen
  \bibfield  {author} {\bibinfo {author} {\bibfnamefont {V.}~\bibnamefont
  {Mustonen}}\ and\ \bibinfo {author} {\bibfnamefont {M.}~\bibnamefont
  {Lässig}},\ }\href {\doibase 10.1016/j.tig.2009.01.002} {\bibfield
  {journal} {\bibinfo  {journal} {Trends in Genetics}\ }\textbf {\bibinfo
  {volume} {25}},\ \bibinfo {pages} {111} (\bibinfo {year} {2009})}\BibitemShut
  {NoStop}%
\bibitem [{\citenamefont {Mustonen}\ and\ \citenamefont
  {Lassig}(2010)}]{mustonenFitnessFluxUbiquity2010}%
  \BibitemOpen
  \bibfield  {author} {\bibinfo {author} {\bibfnamefont {V.}~\bibnamefont
  {Mustonen}}\ and\ \bibinfo {author} {\bibfnamefont {M.}~\bibnamefont
  {Lassig}},\ }\href {\doibase 10.1073/pnas.0907953107} {\bibfield  {journal}
  {\bibinfo  {journal} {Proceedings of the National Academy of Sciences}\
  }\textbf {\bibinfo {volume} {107}},\ \bibinfo {pages} {4248} (\bibinfo {year}
  {2010})}\BibitemShut {NoStop}%
\bibitem [{\citenamefont
  {Kampen}(2007)}]{kampenStochasticProcessesPhysics2007}%
  \BibitemOpen
  \bibfield  {author} {\bibinfo {author} {\bibfnamefont {N.~G.~v.}\
  \bibnamefont {Kampen}},\ }\href@noop {} {\emph {\bibinfo {title} {Stochastic
  processes in physics and chemistry}}},\ \bibinfo {edition} {3rd}\ ed.,\
  North-{Holland} personal library\ (\bibinfo  {publisher} {Elsevier},\
  \bibinfo {address} {Amsterdam ; Boston},\ \bibinfo {year} {2007})\ \bibinfo
  {note} {oCLC: ocm81453662}\BibitemShut {NoStop}%
\end{thebibliography}
\end{document}